\documentclass{aa}
\usepackage{graphicx}
\usepackage{hyperref}
\usepackage{txfonts}
\usepackage{xcolor}
\usepackage{caption}
\usepackage{subcaption}
\usepackage{textcomp}
\usepackage[rightcaption]{sidecap}
\graphicspath{{Figures/}}

\begin{document}

\title{Coronal dimmings as indicators of early CME propagation direction}

\author{Shantanu Jain\inst{1}
       and Tatiana Podladchikova \inst{1}
       and Galina Chikunova\inst{1}
       and Karin Dissauer\inst{2} 
       and Astrid M. Veronig\inst{3,4}
}

\institute{Skolkovo Institute of Science and Technology, Bolshoy Boulevard 30, bld. 1, 121205, Moscow, Russia \\
\email{Shantanu.Jain@skoltech.ru}
\and NorthWest Research Associates, 3380 Mitchell Lane, Boulder, CO 80301, USA
\and
University of Graz, Institute of Physics, Universit\"atsplatz 5, 8010 Graz, Austria\
\and
University of Graz, Kanzelh\"ohe Observatory for Solar and Environmental Research, Kanzelh\"ohe 19, 9521 Treffen, Austria
}

\date{Received September, 2023; accepted November, 2023}

\abstract
{Coronal mass ejections (CMEs) are large-scale eruptions of plasma and magnetic field from the Sun that can cause severe disturbances in Space Weather. Earth-directed CMEs are responsible for the disruption of technological systems and damaging power grids. However, the early evolution of CMEs, especially Earth-directed ones, is poorly tracked using traditional coronagraphs along the Sun-Earth line.}
{The most distinct phenomena associated with CMEs in the low corona are coronal dimmings that are localized regions of reduced emission in the extreme-ultraviolet (EUV) and soft X-rays, formed due to mass loss and expansion during a CME. We present a new approach to estimate the early CME propagation direction based on the expansion of coronal dimmings.}
{We develop a method called DIRECD (Dimming InfeRred Estimate of CME Direction). First, we perform simulations of CMEs in 3D using a geometric CME cone model and varying parameters such as width, height, source location and deflection from the radial direction to study their influence on the CME projection onto the solar sphere. Second, we estimate the dominant direction of the dimming extent based on the evolution of the dimming area. Third, using the derived dominant direction of the dimming evolution on the solar sphere, we solve an inverse problem to reconstruct an ensemble of CME cones at different heights, widths and deflections from the radial propagation. Finally, we search for which CME parameter combinations, the CME orthogonal projections onto the solar sphere would match the geometry of the dimming at the end of its impulsive phase best, to derive the CME direction in 3D. We test our approach on two case studies on 1 October 2011 and 6 September 2011. We also validate our results with 3D tie-pointing of the CME bubble in EUV low corona and with 3D reconstructions by graduated cylindrical shell modeling (GCS) of white-light CME higher up in the corona.}
{Using DIRECD, we find that the CME on 1 October 2011 expands dominantely towards the South-East, while the CME on 6 September 2011 is inclined towards the North-West. This is in agreement with the CME direction estimates from previous studies using multi-viewpoint coronagraphic observations.}
{Our study demonstrates that coronal dimming information can be used to estimate the CMEs direction early in its evolution. This allows us to provide information on the CME direction early on before it is observed in the coronograph's field-of-view, which is of practical importance for space weather forecasting and the mitigation of potential adverse impacts on Earth.}


\authorrunning{S. Jain et al.}
\maketitle

\section{Introduction} 
Coronal Mass Ejections (CMEs) are the most powerful events in the solar system that cause severe disturbances of our space weather. They are clouds of magnetized plasma that are expelled from the Sun with speeds of more than $3000~\textrm{km}\,\textrm{s}^{-1}$ \citep{gomez2020clustering,michalek2009expansion, tsurutani2014extreme, gopalswamy2009soho,Veronig2018_Genesis}. CMEs can be harmful to technological systems as they can, e.g., affect radio transmission, induce currents in power grid systems and lower the orbits of satellites orbiting the Earth \citep{sandford1999impact,doherty2004space,baker2013major}. Thus, to have proper lead times, it is important to detect and study already the early evolution of CMEs, in particular Earth-directed ones. Since the early CME evolution is not well observed with coronagraphs and since Earth-directed CMEs are prone to strong projection effects \citep{schwenn2005association}, it is important to explore whether CME-associated phenomena may provide additional information on the CME properties. 

The most distinct phenomena associated with CMEs are coronal dimmings, which are manifested as regions of reduced extreme ultraviolet (EUV) and X-ray emissions in the low corona \citep{hudson1996long,sterling1997yohkoh,thompson1998soho}. This sudden reduction in the coronal emission is attributed to the expansion of the CME structure resulting in density and mass depletion. This interpretation is supported by differential emission measure (DEM) studies showing substantial density drops in dimming regions \citep[e.g.][]{ cheng2012differential, vanninathan2018plasma,veronig2019spectroscopy} and spectroscopic observations of strong outflows \citep[e.g][]{harra2001material, jin2009coronal,tian2012can}. The close relationship between the early evolution of CMEs and coronal dimmings in the low corona has been investigated in a number of studies. There have been several case studies analyzing the relation between coronal dimmings and CME mass (e.g. \citep{harrison2000spectroscopic, harrison2003coronal, zhukov2004nature, lopez2017mass, wang2023observations} showing strong positive correlation, the morphology and early evolution of the CME \citep{attrill2006using, qiu2017gradual}, and their timing \citep{miklenic2011coronal}. In addition, studies have also investigated the statistical relationship between CMEs and coronal dimmings \citep{bewsher2008relationship, reinard2009relationship,mason2016relationship,
krista2017statistical, aschwanden2017global, dissauer2018, dissauer2019statistics}. It was found that the CME mass shows a high correlation with dimming area, dimming brightness and the magnetic flux covered by the dimming region, while the maximum speed of the CME is strongly correlated with the corresponding time derivatives (i.e., area growth rate, brightness change rate, and magnetic flux rate) as well as the mean intensity of the dimmings \citep{dissauer2018statistics, dissauer2019statistics, chikunova2020coronal}. 

Since CMEs influence the space weather around Earth and other planets, an understanding of the CME trajectory is of the utmost importance. There have been several attempts to determine the propagation direction of CMEs and their deflections \citep{hildner1977mass,macqueen1986propagation} using coronagraph measurements. With the launch of the STEREO mission and the possibility to simultaneously observe CMEs from multiple viewpoints in the heliosphere, a variety of methods have been developed and applied to reconstruct the 3D direction and geometry of CMEs \citep{thernisien2006modeling,mierla20103,isavnin2013three,liewer2015observations, kay2017deflection}. Several studies have tried to model the CME propagation and deflection using a graduated cylindrical shell (GCS) model \citep{thernisien2006modeling, thernisien2011implementation} by varying certain CME parameters such as width, height, tilt, latitude, longitude etc. \citep{gui2011quantitative, kay2017deflection, temmer2017flare}. \cite{byrne2010propagation} used an elliptical tie-pointing method to create a 3D model of CMEs and found that the deflections obtained for the events under study were consistent with the in-situ observations. \cite{liu2010reconstructing} reconstructed 3D models of several events using a geometric triangulation method and showed that this method can link solar observations with corresponding in-situ signatures at 1~AU and predict CME arrival at Earth. \cite{lugaz2011numerical,lugaz2012deflection} studied CME deflection using magneto-hydrodynamic models (MHD) and found no strong rotation of 22 August 2005 CME as it propagated. However, 23-24 May 2010 CMEs showed a strong deflection after collision.

\cite{zuccarello2011role,Kilpua2009STEREOOO} showed that during solar minima, CMEs originating from high latitude can be easily deflected towards the the heliospheric current sheet (HCS) as a result of the influence of the magnetic field within a coronal hole (CH), eventually resulting in geoeffective events. This phenomenon primarily occurs during periods of solar minimum and applies specifically to CMEs with very slow velocities. These slower CMEs encounter challenges in overcoming the restraining influence of the magnetic field above them. Consequently, they tend to be steered by the polar coronal magnetic fields, leading to in-situ effects occurring in close proximity to the ecliptic plane.
\cite{gopalswamy2009cme} showed that the coronal holes may have deflected the associated coronal mass ejections away from the Sun-Earth line during the declining phase of solar cycle 23.
\cite{sahade2020influence} established a relation between CH properties, the location of the minimum magnetic field region and CME deflection to show that the minimum magnetic energy region, responsible for the deflection, is associated with the presence of the CH near the CME. 
In addition, CMEs can also be channeled by the active region coronal magnetic field itself \citep{mostl2015strong,wang2015role}. 
\cite{shen2011kinematic} showed evidence that the trajectory of the CME was influenced by the background magnetic field, and the CME tends to propagate towards the region with lower magnetic energy density. \cite{gui2011quantitative} extended the work with additional observations and showed that the CME deflection is consistent with the direction of the background gradients. 
ForeCAT developed by \citep{kay2013forecasting} serves to forecast CME deflections using magnetic forces and CME rotations \citep{kay2015global}. Using ForeCAT, \cite{kay2016using} found that CME deflections for 8 April 2010 and 12 July 2012 events. \cite{mierla20103} compared different CME reconstruction techniques to find CME properties like speed, mass and direction of propagation. The different reconstruction techniques estimated the most probable CME propagation direction at the outer boundary of the corona (observed in coronagraph images) to be within $10^{\circ}$ of each other.
The relationship between coronal dimmings, CME propagation and deflection has not been directly investigated in the literature yet, although there exist several studies that highlight its importance. \cite{thompson2000coronal} analyzed seven CMEs that occurred between 23 April and 9 May 1998, and found that the extended dimming areas in these events generally mapped out the apparent ``footprint'' of the CME as observed by white-light coronagraphs. Some studies have also tracked the CME and high speed solar wind (HSS) interaction \citep{heinemann2019cme} and CME-shock interactions \citep{mostl2015strong}. Both studies show the dimming evolution and study the direction. In a recent study, \citet{Chikunova2023} demonstrated that the overall dimming morphology closely reflects the inner part of the 3D GCS reconstruction of a CME, linking the 2D dimming to the 3D CME bubble and suggesting the use of dimming observations to obtain insight into the CME direction. The SunCET mission \citep{mason2021}, dedicated to the study of CME acceleration processes, suggests using coronal dimmings as a source of information on the Earth-directed CMEs. 

In this study we present a new method called DIRECD (Dimming InfeRred Estimate of CME Direction) to use the extent and morphology of coronal dimmings to estimate the early 3D direction of the corresponding CME.
We test our approach on case studies of two well-observed dimmings and CMEs on 1 October 2011 and 6 September 2011. Note, that in this study we focus on dimmings due to the expansion and evacuation of mass by the erupting CME, but not on more shallow dimmings interpreted as the rarefaction regions propagating behind EUV waves \citep{Muhr2011,Vrsnak2016,Veronig2018_Genesis,Podladchikova2019three}. 

\begin{figure}
\begin{subfigure}{\columnwidth}
  \centering
  \includegraphics[width=0.65\linewidth]{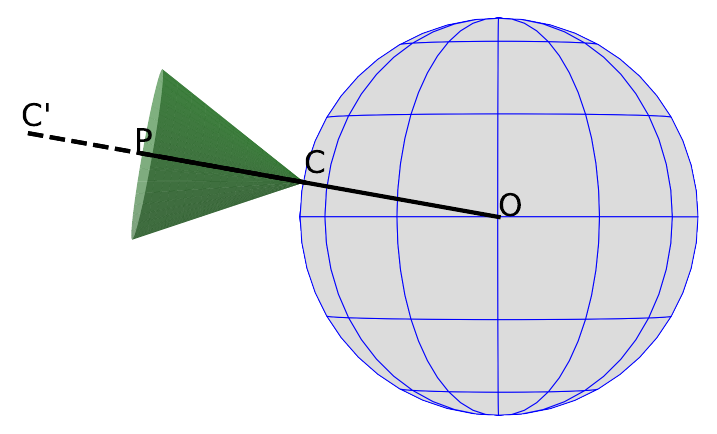}
  \caption{}
  \label{fig:3D_radial}  
  \includegraphics[width=0.65\linewidth]{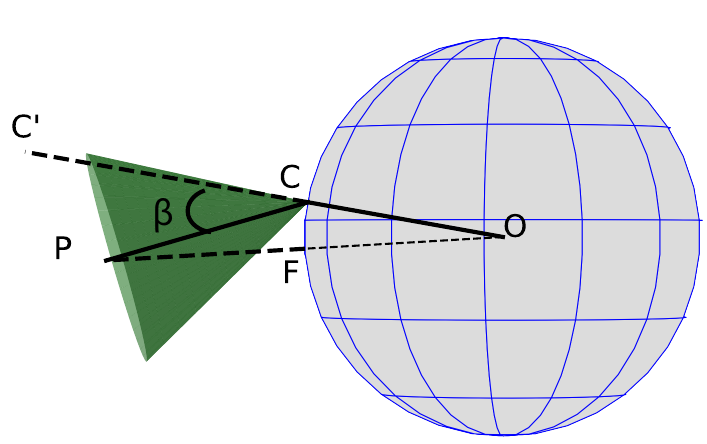}
  \caption{}
  \label{fig:3D_inclined}  
  \end{subfigure}
\caption{CME cone model. Panel~(a) shows the cone in the radial direction with angular width of $\phi=28^{\circ}$, while panel~(b) represents the same cone but inclined from the radial direction by an angle $\beta=25.52^{\circ}$. Point C is the source region on the solar sphere, P is the top point of the cone central axis, O is the center of the Sun, and F is the orthogonal projection of the point P onto the solar sphere.} 
\label{CME_cone_3D}
\end{figure}

\begin{figure*}  
	\centering	\includegraphics[width=0.67\textwidth] {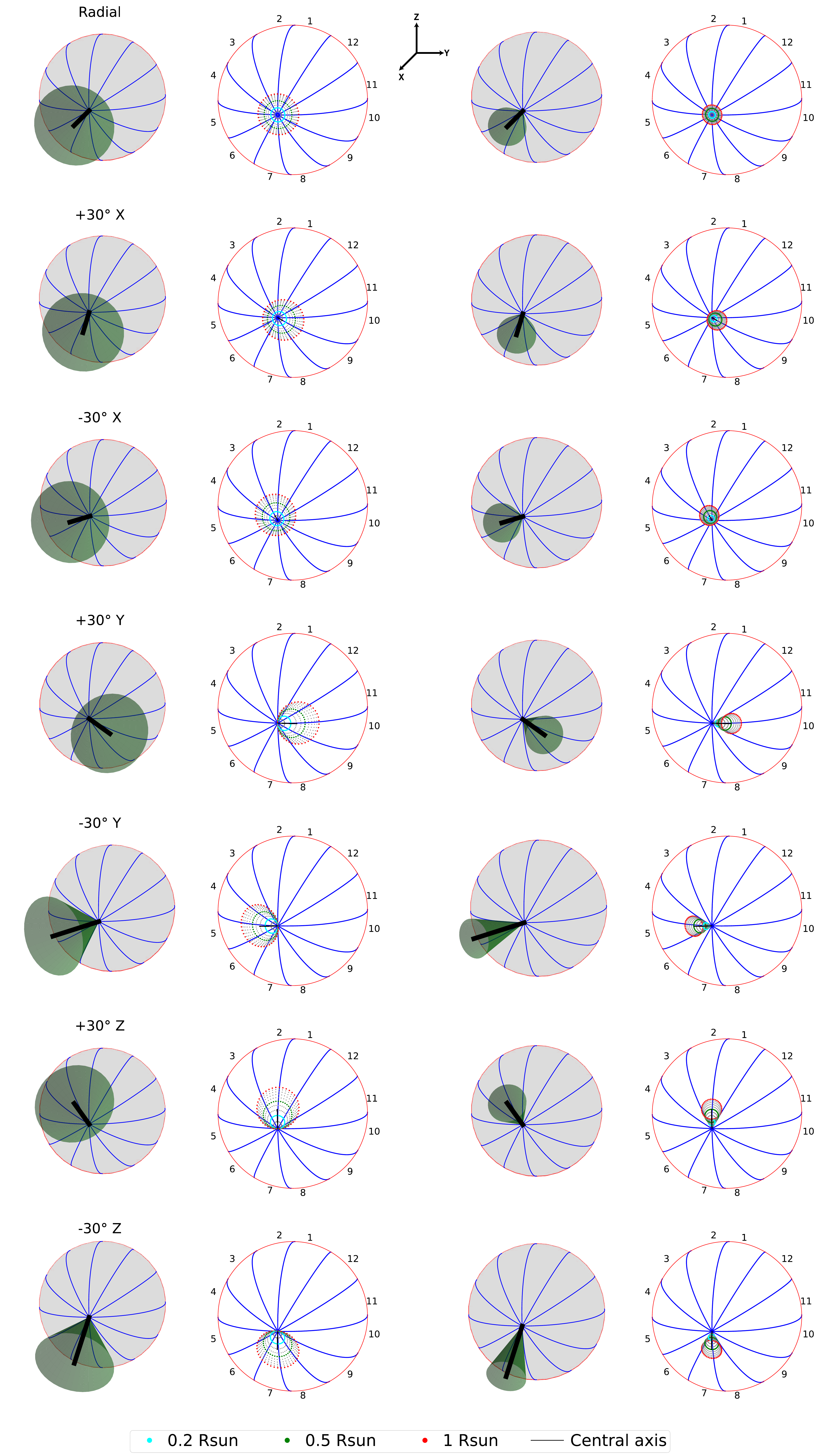}
	\caption{Simulation of the CME cone and its orthogonal projection in quadrant 3 (near center). From left to right: the CME cone with a $30^{\circ}$ of angular width (column 1) and its orthogonal projection (column 2) on the solar sphere; the CME cone with a $15^{\circ}$ of angular width (column 3) and its orthogonal projection (column 4). From top to bottom: first row - the radial direction, second and third rows show the case of $\pm$ 30$^{\circ}$  from inclination from the X-axis  direction in the YZ plane, fourth and fifth rows $\pm$ 30$^{\circ}$  inclination from the Y-axis direction, and sixth and seventh rows $\pm$ 30$^{\circ}$ inclination from the Z-axis direction. The orthogonal projections at 0.2 $R_{sun}$, 0.5 $R_{sun}$ and 1 $R_{sun}$ are shown by cyan, green and red markers. The grey markers show the projections of intermediate heights from 0.1-1 $R_{sun}$.
 } 
	\label{sim_3_nl}
\end{figure*}

\section{Methods and Analysis}\label{Methods}
Our method is based on certain assumptions regarding the relationship between dimmings and CMEs, as well as their key parameters influencing this connection. Several important parameters influence the correlation between dimmings and CMEs, including the height at which the CME remains connected to the dimming, the CME width, the inclination from the radial propagation, as well as the source location. In our analysis, we consider the dimming at the end of the impulsive phase i.e., time range within the majority of the dimming region develops (see \citep{dissauer2018}), when the connections between an expanding CME, leaving imprints in solar corona and dimming are fully formed.

Our approach to identify the early propagation of CMEs from the expansion of dimmings involves the following main steps. First, we simulate CMEs in 3D using a geometric cone model to establish the relation between the 3D CME and the 2D projection on the solar disk in the form of the dimming. We investigate how the orthogonal projections of the cone onto the solar sphere (which can be considered as dimming signatures) depend on various sets of CME parameters, such as width, height, source location, and inclination from the radial propagation (Section~\ref{CME_simulations}).

Secondly, we determine the dominant direction of dimming development by observing how the dimming area evolves (Section~\ref{Dimming_direction}). Thirdly, using the identified dominant direction of dimming evolution on the solar sphere, we address an inverse problem to reconstruct a set of CME cones at various heights, widths, and deviations from radial propagation. Then, we search for the CME parameters, for which the CME orthogonal projections onto the solar sphere would match the geometry of the dimming at the end of its impulsive phase. Finally, we determine the resulting 3D direction of the CME, including also the direction with respect to the inclination from the radial direction in both the East-West and North-South planes   (Section~\ref{Dimming_CME_direction}). We utilize the DIRECD method in our analysis, focusing on the event that occurred on October 1, 2011 (Section~\ref{Dimming_direction}~and~\ref{Dimming_CME_direction}) and compare them with the 3D reconstructions obtained from STEREO/EUVI images (Section~\ref{3D_CME}). Additionally, we apply this approach to examine the fast event that occurred on September 6, 2011 (Section~\ref{case_study_2}). 

\subsection{Simulations with a CME cone model}\label{CME_simulations}
CMEs occur in many different shapes and thus constructing CME models that can fit the observations is the first important step in any CME direction study. In order to relate the 2D on-disk dimming signature to the 3D CME structure, we model the CME as a cone \citep{zhao2002determination,xie2004cone,mays2015ensemble} with certain direction, height and width, and derive the orthogonal projections onto the solar sphere which we take as an estimate how the plasma evacuation would be manifested as a coronal dimming when projected onto the solar sphere.

Figure~\ref{CME_cone_3D} shows examples of CME cones in 3D, where panel~(a) shows the case where the CME cone propagates in radial direction along the line OC\textquotesingle~and has an angular width of $\phi=28^{\circ}$. Here, point O indicates the center of the Sun, and points C (source region on the solar sphere) and P (top point of the cone central axis) belongs to the line OC\textquotesingle. Panel~(b) shows the cone along CP, which is inclined with respect to the radial direction OC\textquotesingle~by an angle $\beta=25^{\circ}$. Throughout the text, we call the angle $\beta$ ``inclination angle''. 

We model a CME cone for various combinations of CME parameters such as CME source location, width and inclination angle, and derive their orthogonal projections onto the solar sphere as an estimate how the coronal dimming shape might look like. We show Simulations at nine different source region locations, where one is in the disk center, and the other 8 are in 4 quadrants (2 points per quadrant: one near the disk center and the other one near the limb). Throughout the text, by quadrant 1 we mean the top-right (NW) part of the solar sphere, quadrant 2 -- top-left (NE), quadrant 3 -- bottom-left (SE), and quadrant 4 -- bottom-right (SW).

Figure~\ref{sim_3_nl} shows a CME simulation in quadrant 1 (near center) as an example. The sphere and cone is constructed in such a way that the vertical direction corresponds to the Z-axis (solar north/south), the horizontal one corresponds to the Y-axis (solar east/west), and the X-axis is perpendicular to both these axes (into/out of the page). The CME is modeled by using two cones of different angular widths ($15^{\circ}$ and $30^{\circ}$) and maximum height of 1~$R_{sun}$ above the solar surface. The first two columns from the left show the CME and the orthogonal projections on to the sphere for $30^{\circ}$ width, whereas the two columns from the right for $15^{\circ}$.

The orthogonal projections of the all cone points are derived using the following relations (and are plotted in figure \ref{sim_3_nl} on the sphere with gray dots representing the CME related dimming expansion):
\begin{equation}\label{ortho}
\begin{array}{cc}
X_{ortho}=\frac{R_{sun}\cdot X_c}{\sqrt{(X_c)^2+(Y_c)^2+(Z_c)^2}} \\
Y_{ortho}=\frac{R_{sun}\cdot Y_c}{\sqrt{(X_c)^2+(Y_c)^2+(Z_c)^2}} \\ 
Z_{ortho}=\frac{R_{sun}\cdot Z_c}{\sqrt{(X_c)^2+(Y_c)^2+(Z_c)^2}} \\ 
\end{array}
\end{equation}
Here, $X_{ortho}$, $Y_{ortho}$, $Z_{ortho}$ are the coordinates of orthogonal projections, $X_c$, $Y_c$, $Z_c$ are the coordinates of the cone in 3D (in HEEQ), and  $R_{sun}$ is the radius of the Sun in km. Additionally we show the orthogonal projections at 0.2 $R_{sun}$ (cyan markers), 0.5 $R_{sun}$ (green markers) and 1 $R_{sun}$ (red markers).
The solar sphere is divided into 12 angular sectors of $30^{\circ}$ width each (see more details in Section~\ref{Dimming_direction}).  Black lines indicate the cone central axis (columns 1 and 3) and its orthogonal projection (columns 2 and 4).

As can be seen from the first row (top panel) in Figure~\ref{sim_3_nl}, for the radial cone, the CME projections form concentric circles around the source location of the cone. The area of the CME projection increases with the growth of the CME height and width. The second and third rows in Figure~\ref{sim_3_nl} represent the CME cone inclined at an angle of $\pm$ 30$^{\circ}$ from the X-axis direction in the YZ plane. As it can be seen, the projections are no longer concentric circles but ellipses with the largest extension at sector 4 for $+30^{\circ}$ and sector 9 for $-30^{\circ}$ inclination. For the inclinations in the Y-axis direction (fourth and fifth rows), the CME projections show the largest extent on the border between sector 10 and 11 for $+30^{\circ}$ and sector 4 and 5 for $-30^{\circ}$. Inclinations in the Z-axis direction (sixth and seventh rows) produce the CME projections mainly located in sectors 1,2 and 12 for $+30^{\circ}$ and in sectors 7,8 and 9 for $-30^{\circ}$.   
Thus, as can be seen from Figure~\ref{sim_3_nl}, the CME height, width and inclination from the radial direction affect the geometry of the CME orthogonal projections. To illustrate the geometry of CME projections in more detail, Figures \ref{sim_cen} -- \ref{sim_4_l} in Appendix~\ref{Appendix_A} show simulations of CME cones originating from other source locations.

\subsection{Determination of dominant dimming expansion}\label{Dimming_direction}
We demonstrate a method on estimating the dominant dimming expansion on a case study of the 1 October 2011 CME event that originated from close to the center of the Sun. \citet{temmer2017flare} analyzed this event to better understand the dynamic evolution of the CME and its embedded magnetic field and tracked the temporal and spatial evolution of the CME in the interplanetary space by reconstructing the CME using the GCS model. The source location of the flare was N10 W06. The authors estimated that the Interplanetary CME (ICME) arrived at Earth after 4 days with an impact speed of 426~$\pm~30 ~ \textrm{km/s}$. Further, the ICME caused a weak geomagnetic storm peak Dst of $-43~nT$. 

\begin{figure}
	\centering
	\includegraphics[width=0.5\textwidth]{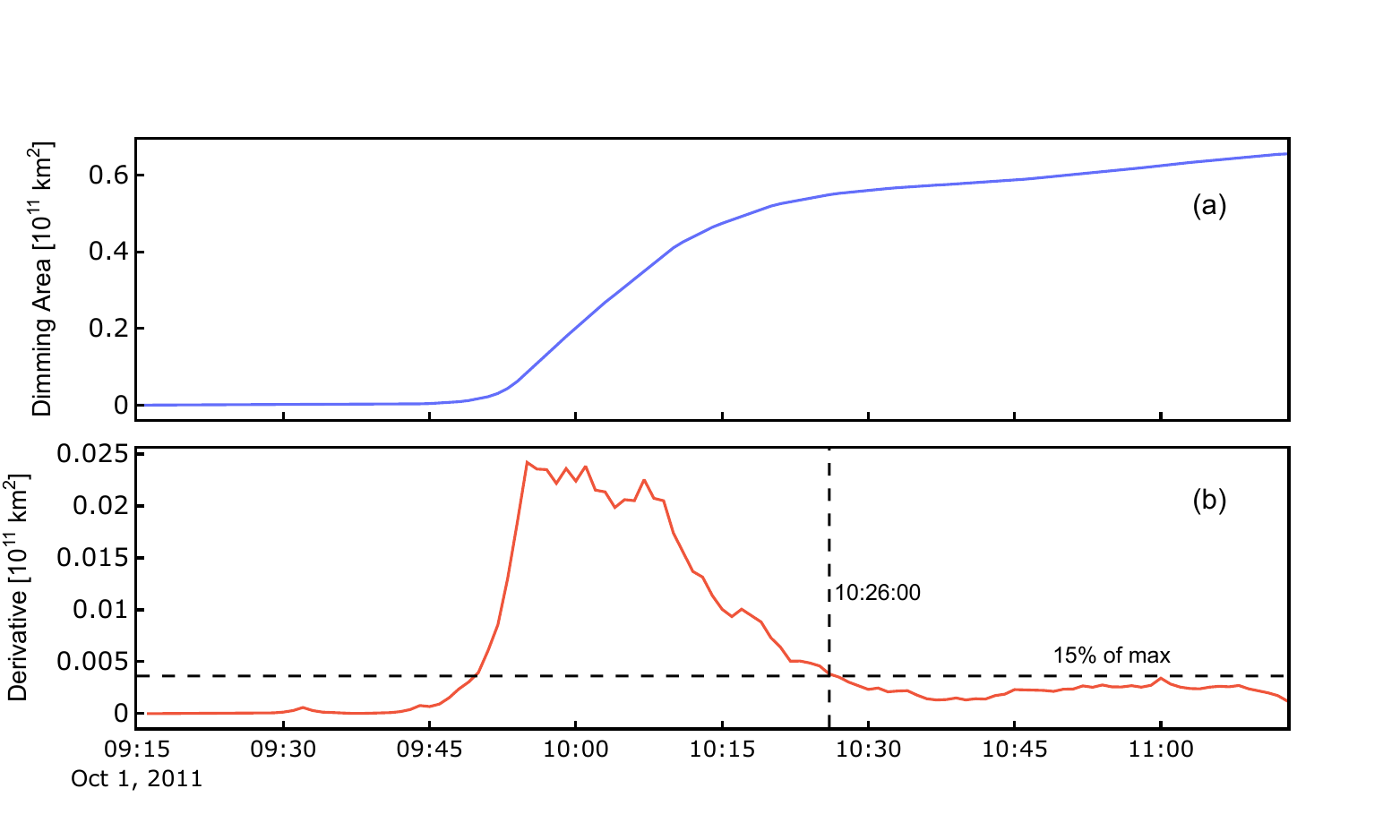}
	\caption{Expansion of dimming area $A_{t}$ (top panel) and its time derivative $dA/dt$ (bottom panel) over 3 hours for the 1 October 2011 event. The end of the impulsive phase is defined as the time when the derivative of the dimming area curve, $dA/dt$ has declined back to 15\% of its maximum value}
\label{area_der_oct2011_max}
\end{figure}

\begin{figure}
	\centering
	\includegraphics[width=0.48\textwidth]{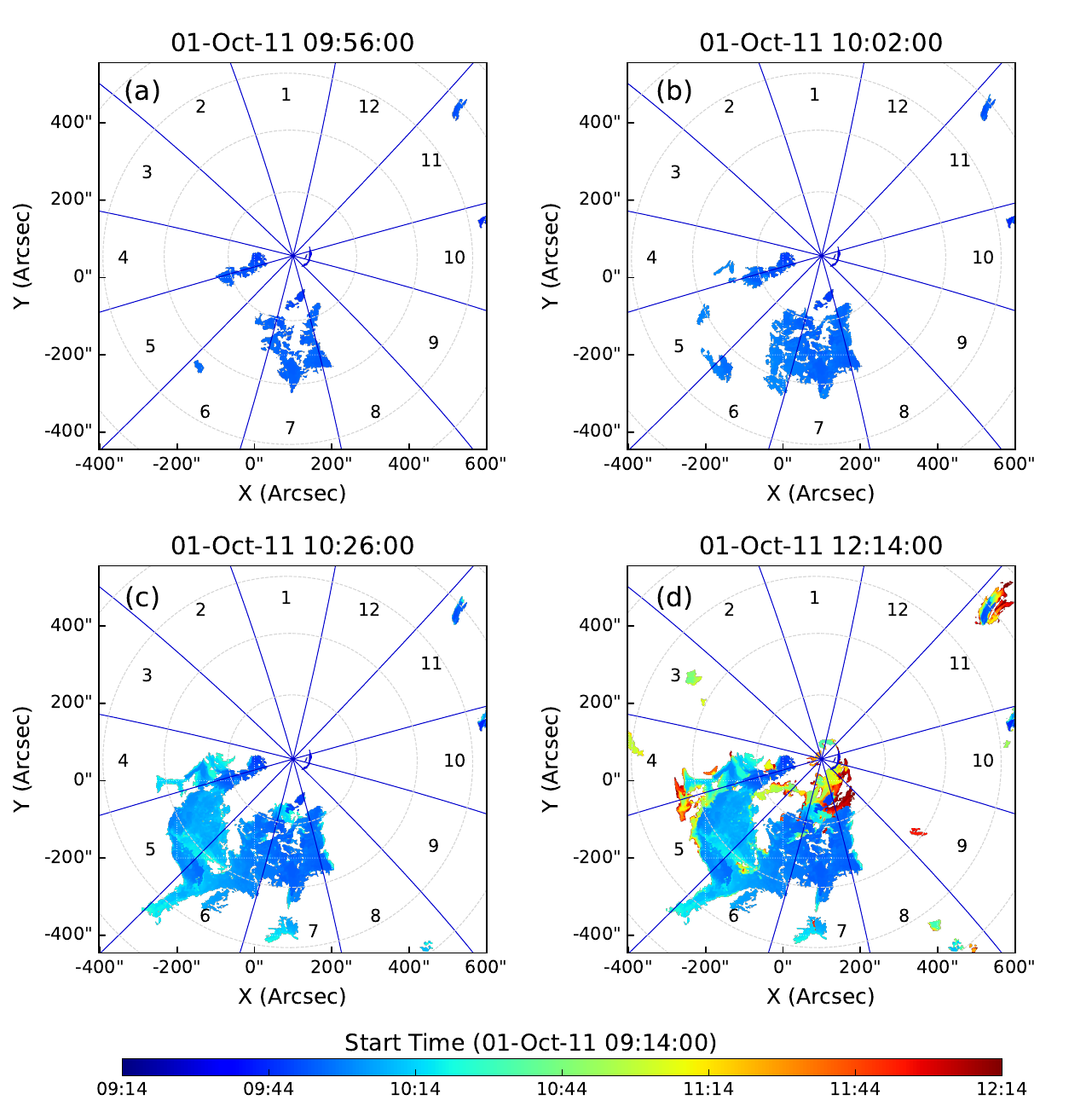}
	\caption{Dimming evolution for the 1 October 2011 CME at (a) the maximum of the impulsive phase (reached 12 minutes after its start), (b) 18~minutes after the start of event, (c) at the end of the impulsive phase (30 minutes after the maximum of the impulsive phase)  and (d) 2.5 hours after the start.  The blue lines indicate 12 angular sectors and the color bar shows when each dimming pixel was detected for the first time.} 
	\label{oct_dimming}
\end{figure}

\begin{figure}
\begin{subfigure}{\columnwidth}
    \centering
    \includegraphics[width=0.7\textwidth] 
  {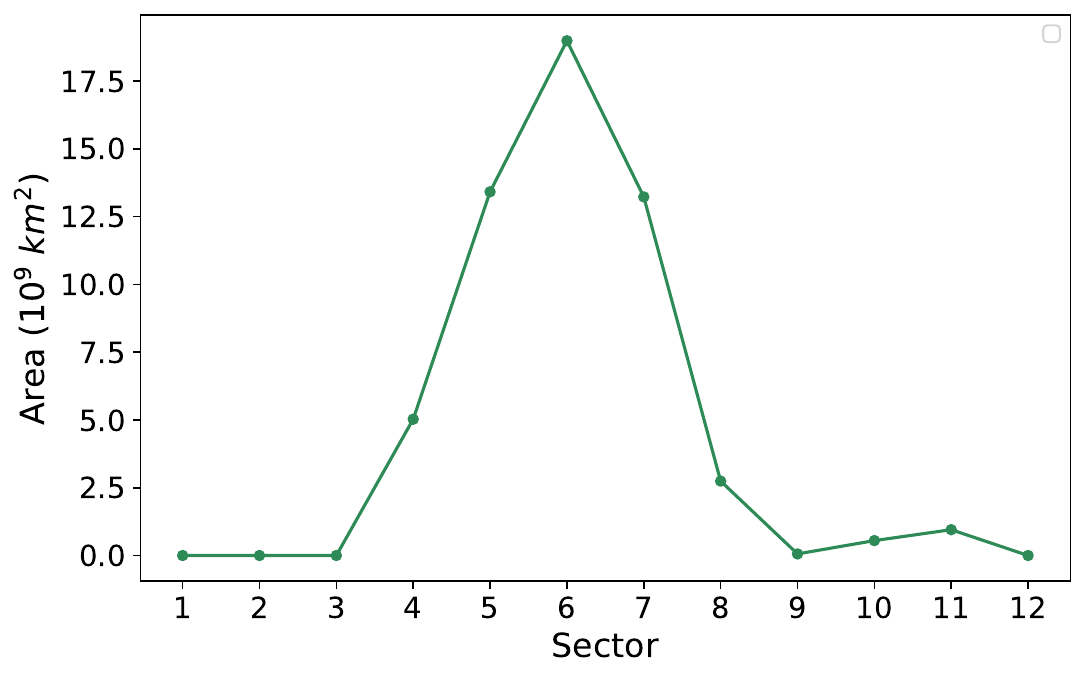}
    \caption{}
    \label{oct_area_sector_end}
\end{subfigure}    
\begin{subfigure}{\columnwidth}    
    \includegraphics[width=0.9\linewidth]
 {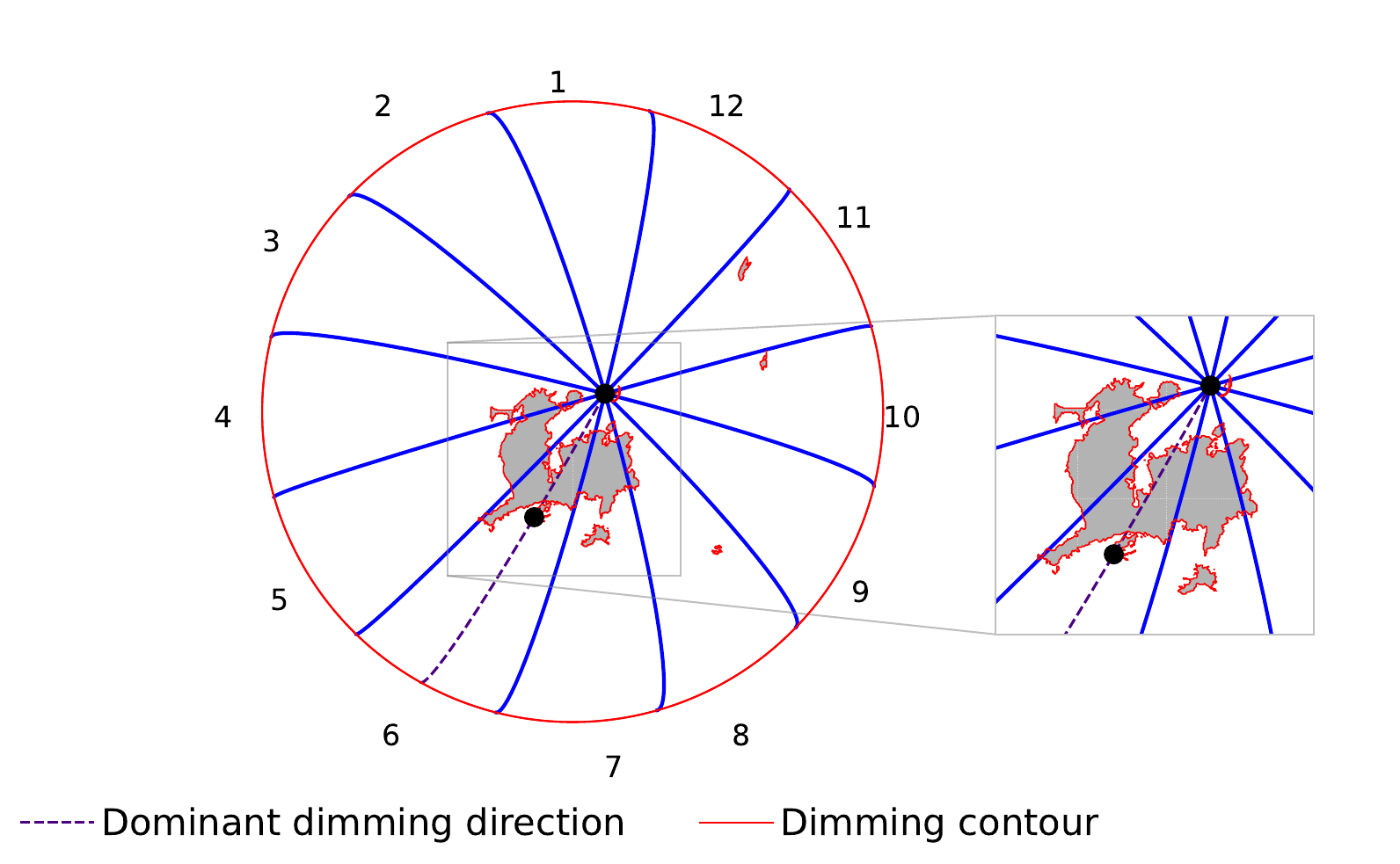}
   \caption{}
    \label{dominant_sector_oct_end}  
\end{subfigure}
\caption{Dominant direction of dimming expansion at the end of impulsive phase for 1 October 2011. Top: Dimming area in the different sectors (shown in the bottom panels), showing its maximum at sector 6.  Bottom: cumulative dimming pixel mask outlined by gray with red contours. The blue lines indicate the 12 sectors: the The dashed purple line indicates the sector of dominant dimming development. 
The black dots show the dimming edges (one of them being the source) in the sector of dominant dimming which are used to generate the CME cones at different heights, associated widths and inclination angles.}
\label{oct_dimming_area_mask}
\end{figure}

To estimate the dominant direction of dimming expansion, we follow the approach introduced in \citet{Chikunova2023}. We first segment the dimming region with an automated detection technique based on thresholding and region growing \citep{dissauer2018statistics,chikunova2020coronal}. The detection algorithm uses a sequence of logarithmic base ratio images from the 211~\AA~ filter of the Atmospheric Imaging Assembly \citep[AIA;][]{Lemen2012}) onboard the Solar Dynamics Observatory \citep[SDO;][]{pesnell2012solar}, from which we derive both instantaneous (for a particular time step) and cumulative (for all time steps within the particular time range) dimming maps. From the dimming area evolution $A(t)$ derived from these maps (Figure~\ref{area_der_oct2011_max}a), we obtain the end time of the impulsive dimming growth. According to \cite{dissauer2018statistics}, this is defined as the time when the cumulative dimming area growth rate $dA/dt$ has decreased to 15\% of its maximum (Figure~\ref{area_der_oct2011_max}b).

Figure~\ref{oct_dimming} shows the evolution of the detected cumulative dimming mask within the first 2.5 hours after the start of the CME on 1 October 2011, 09:44~UTC as observed from coronagraphs. We analyze the dimming extent - at the end (c) of the dimming impulsive phase, when the connections between the dimming and CME are fully formed. Each individual pixel of the presented cumulative dimming pixel mask is colored according to when it was first detected from start time of flare/CME event. Blue regions are identified at the beginning of the event, whereas the red regions are detected later. 

Using the same coordinate system as in Figure~\ref{sim_3_nl} centered at the source region, we split the solar sphere into 12 angular sectors of $30^{\circ}$ width each and derive the dimming area profiles $A(t)$ in each sector. The sectors are numbered in counterclockwise direction with sector 1 pointing towards the North. Figure~\ref{oct_dimming_area_mask} shows the resulting dimming area profile as a function on sector number (a) and the detected cumulative dimming mask (b) for the 1 October 2011 at the end of the impulsive phase (10:26~UT). To achieve a more accurate estimate of the dimming area on the sphere, and in particular close to the solar limb, we use an approach presented in \citet{Chikunova2023} that allows to estimate the surface area of a sphere for every pixel\footnote{{\url{https://github.com/Chigaga/area_calculation}}}. As can be seen in Figure~\ref{oct_dimming_area_mask}, the dominant dimming direction is defined by the sector of largest cumulative dimming area extent, which is in this case sector 6.

\begin{figure}
	\centering
	\includegraphics[width=0.35\textwidth]{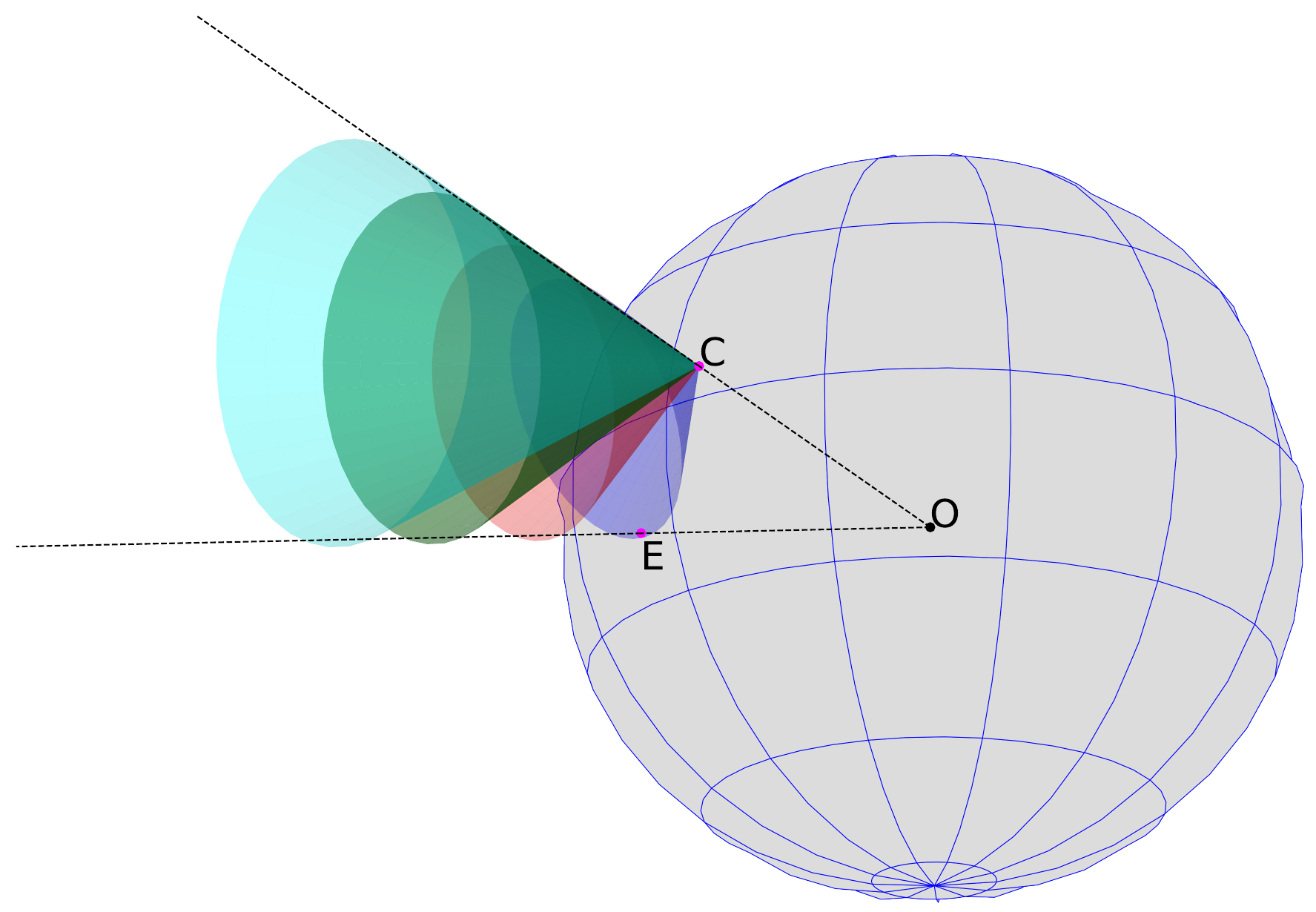}
	\caption{Ensemble of cones of different heights, widths and inclination angle bounded by the dimming edges C and E. The dotted lines outline the dimming edges extended into space connected through the Sun center.
 } 
	\label{ensemble_cones}
\end{figure}

\subsection{Connecting the dimming and CME propagation direction}\label{Dimming_CME_direction}
Once the dominant dimming direction is determined, we create an ensemble of CME cones that originate at the source location and extend into 3D space such that the edges of the cone's orthogonal projections remain confined within the dimming edges (see Figure 5b). We first construct two 3D lines, where one line connects the center of the Sun and the source location, and the second line goes through the center of the Sun and the edge of the dimming along the dominant dimming direction (Figure~\ref{ensemble_cones}, black lines). In case the dimming is developing around the source region, the edge of the dimming that is opposite to the dominant dimming direction could be chosen instead of the source location. We then generate an ensemble of cones in 3D, where the edges are confined by these 3D lines (Figure~\ref{ensemble_cones}, colored cones). This enables the CME cone projections to fit the dimming geometry along the direction of dominant dimming evolution. Moreover, each CME cone in this ensemble will have a unique height (measured perpendicular to the solar sphere) at which the CME remains connected to the dimming and which is uniquely associated with a certain width and inclination angle. 

\begin{figure}
	\centering	\includegraphics[width=0.5\textwidth]{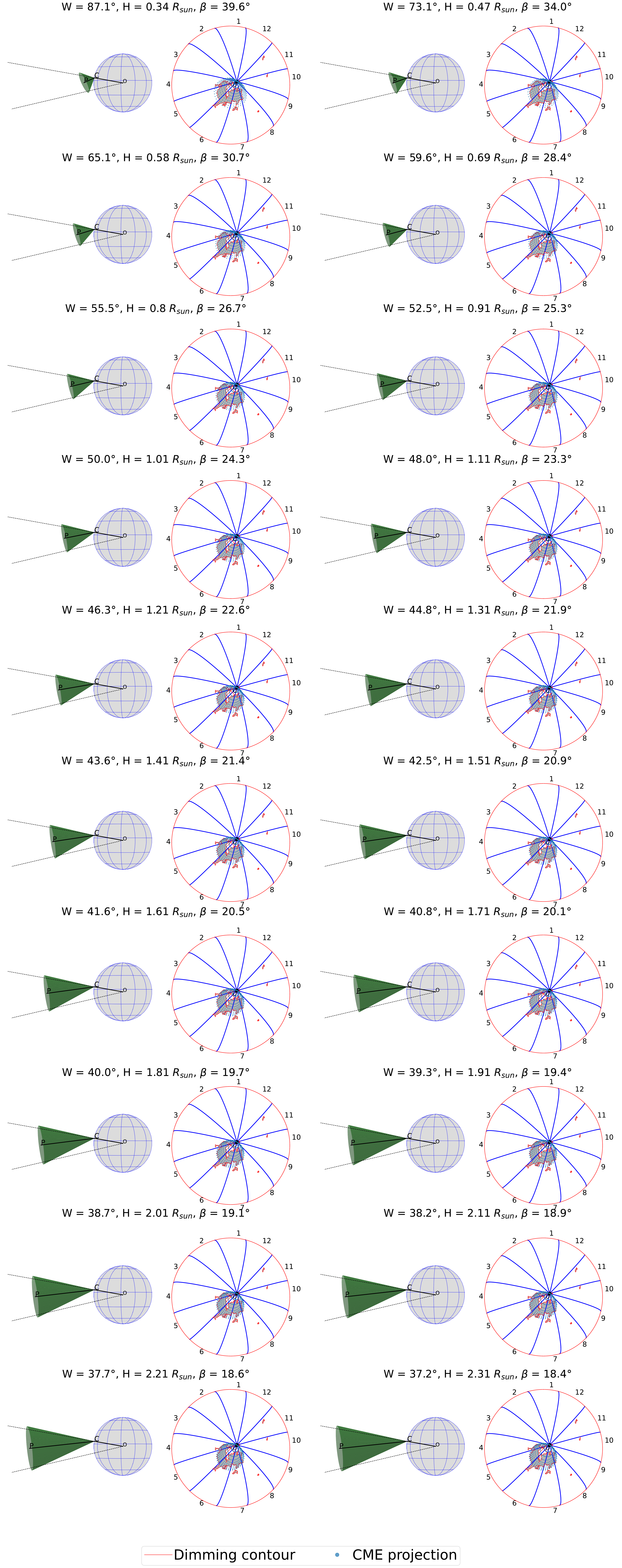}
	\caption{CME cones at heights of 0.27--2.23~$R_{sun}$ for 1 October 2011 (columns 1 and 3) with associated widths and inclination angles and their orthogonal projections onto the solar sphere (columns 2 and 4) bounded by the dimming edges. Columns 1 and 3 depict the side view to better show the reconstructed cones, columns 2 and 4 show the face-on view.
 }
	\label{oct_proj_end}
\end{figure}

\begin{figure}
	\centering	\includegraphics[width=0.48\textwidth]{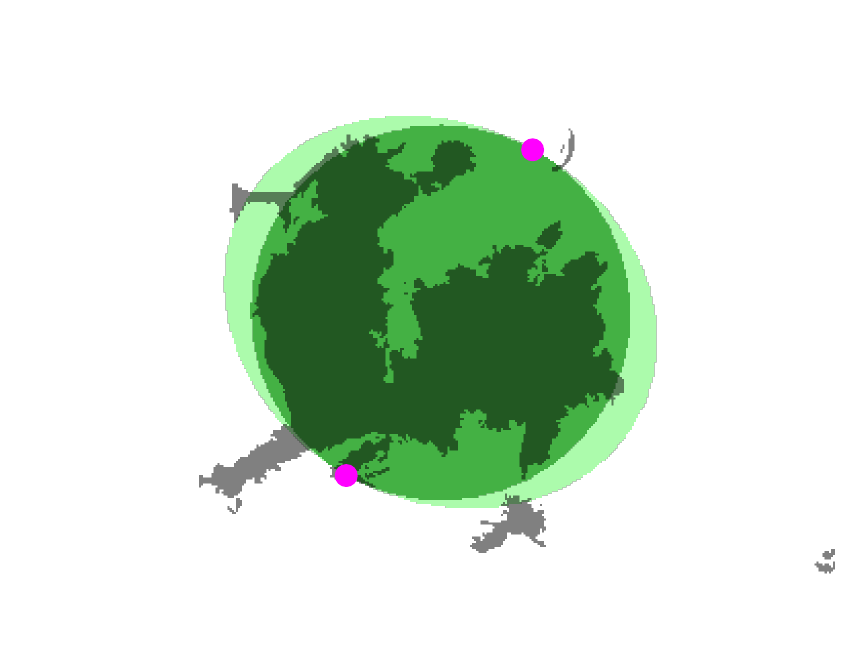}
	\caption{Projections of cone at the smallest height (0.27~$R_{sun}$, light green) and at the largest height (2.82~$R_{sun}$, dark green) together with the dimming mask (gray). The smaller the cone height, the wider the projection. Magenta dots indicate the source location and dimming edge along the direction of dominant dimming evolution.} 
	\label{projection_width}
\end{figure}

Figure~\ref{oct_proj_end} shows twenty generated CME cones at heights over a range of 0.27--2.23~$R_{sun}$, widths of 37.4--91.5$^\circ$ and inclination angles of 18.7--45.7$^\circ$ (columns 1 and 3) and their orthogonal projections onto the solar sphere (columns 2 and 4), which are confined within the dimming edges in the sector of dominant dimming direction at the end of its impulsive phase for 1 October 2011. Columns 1 and 3 are rotated by $90^{\circ}$ for better visibility of the cone evolution such that the Y-axis points to the observer and the X and Z-axes are along the horizontal and vertical directions. Columns 2 and 4 show the face-on view such that X-axis points to the observer, Y and Z-axes are along the horizontal and vertical directions respectively. As can be seen from Figure~\ref{oct_proj_end}, the cone projections cover the dimming, however with an increase of height, the cone projections become narrower. The cone projections take the shape of an ellipse with the minor axis (connecting the source location and the dimming edge along the direction of dominant dimming evolution) being the same for all the cones. At the same time, the major axis of the ellipse is shrinking with increase of cone height, which makes the projection to approach a circular shape due to a smaller inclination angle (less deformation). Figure~\ref{projection_width} illustrates the difference in projection geometry over the lowest height (0.27~$R_{sun}$, light green) to the largest height (2.82~$R_{sun}$, dark green). While the projections do not change along the the direction of dominant dimming evolution (marked by magenta dots) by definition, the sides of projections are the narrowest for the largest height of 2.82~$R_{sun}$. 

\begin{figure}
	\centering
\includegraphics[width=0.48\textwidth]{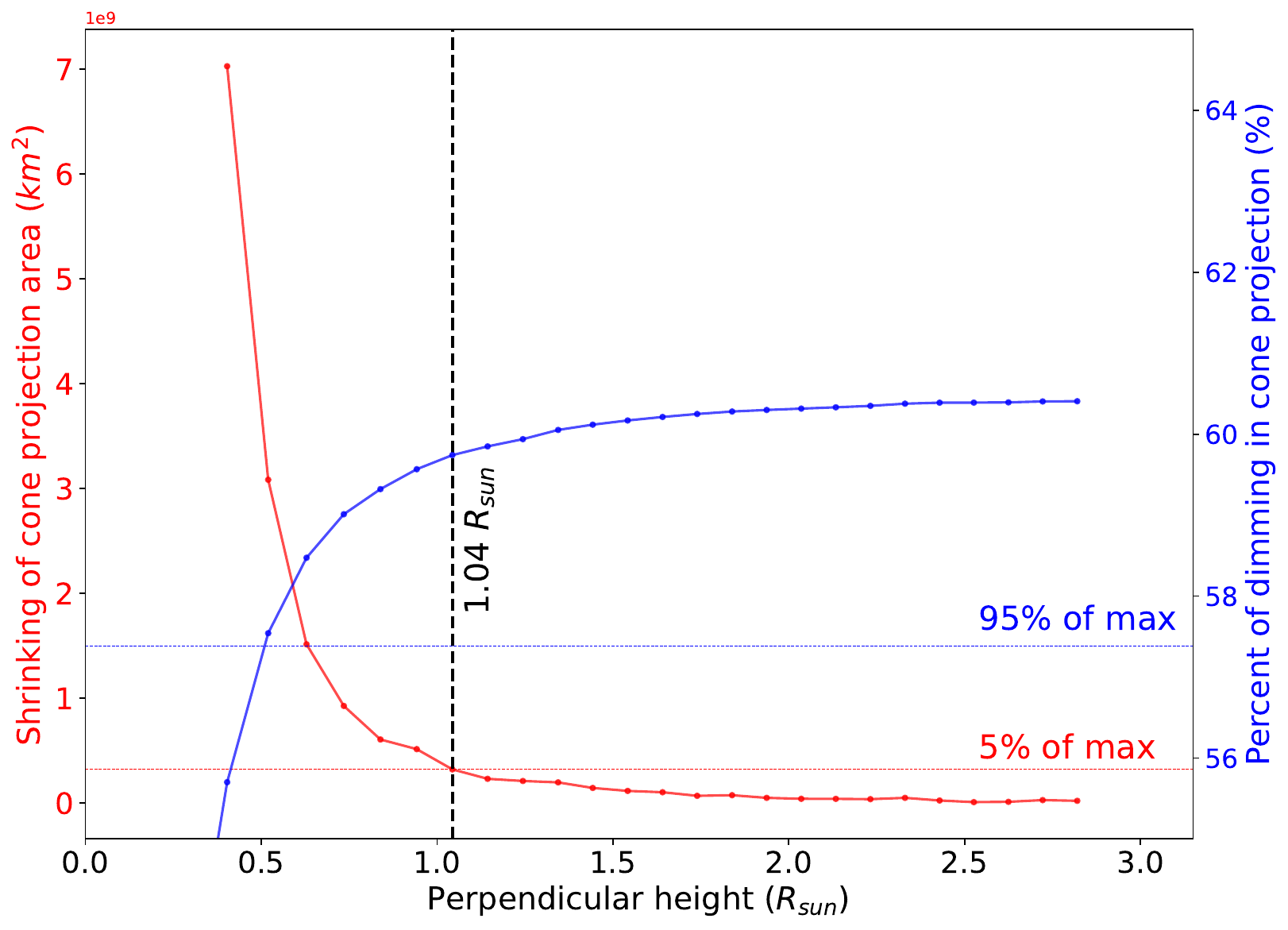}
	\caption{Consecutive differences in projection area of the generated cones ensemble(red, left Y-axis) and percentage of dimming area in projection to the total projection area (blue, right Y-axis) at the end of the impulsive phase of the 1 October 2011 event. Vertical dashed line indicates the step (associated cone height of 1.04~$R_{sun}$, width of 48.7$^\circ$ and inclination angle of 24.3$^\circ$), where consecutive differences differences reach 5\% of the maximum of the cone projection area, which also lies within 95\% of the maximum percentage of dimming area in projection.} 
	\label{dim_area_oct2011}
\end{figure}

At the same time, as can be seen from Figure~\ref{oct_proj_end}, the intense shrinking of the cone projections happens at lower heights from around 0.27--1~$R_{sun}$, while at larger heights the changes are already insignificant. The dynamics of these changes can be quantified by subtracting the projection area of a cone with a smaller height (wider projection) from a projection area of a corresponding cone with a larger height (narrower projection), and the results are shown in Figure~\ref{dim_area_oct2011} (red line, left Y-axis). We chose the cone with best fit to the dimming at the end of the most intense shrinking of projection areas, where the cone projections are still sensitive to the changes in cone parameters, which we estimate at 95\% (vertical dashed line) of the red line peak. In the current case, this corresponds to the cone with a height of 1.04~$R_{sun}$, width of 48.7$^\circ$ and inclination angle of 24.3$^\circ$. Additionally we require that the majority of the dimming area is inside the cone projection and also derive the percentage of dimming area inside the cone projection (blue line, right Y-axis). We accept the solution, if it lies within 95\% of the blue line peak. The chosen empirical thresholds will be validated in future statistical studies. Note, if from other sources we know just one of the CME parameters (height, width, or inclination angle), we could immediately estimate the whole set of parameters and define the CME direction, as each cone in the given ensemble of cones confined by the dimming edges has a height that is uniquely associated with a certain width and inclination angle.

\begin{figure}
	\centering	\includegraphics[width=0.48\textwidth]{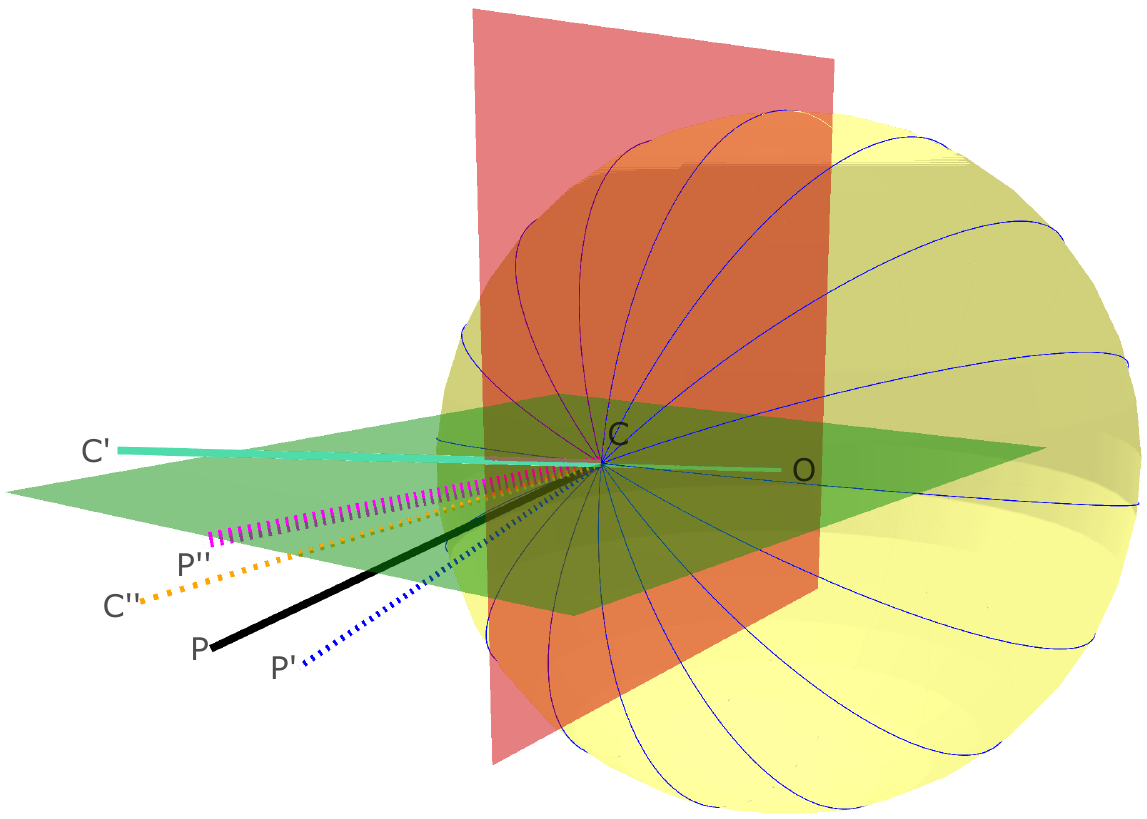}
	\caption{Projection of the CME direction onto the meridional (red) and equatorial (green) planes. The radial direction along the line CC\textquotesingle~ lies in the red meridional plane by definition, the dashed orange line shows its projection onto the equatorial plane. The dashed blue line gives the projection of the cone central axis CP (the CME direction in 3D) to the meridional plane, and the dashed magenta line to the equatorial plane.} 
	\label{plane_oct_end}
\end{figure}

Additionally, we project the resulting CME direction onto the meridional and equatorial planes, and derive the 2D angles between the projections in each plane. Figure~\ref{plane_oct_end} shows the central CME cone axis CP with a height of 1.04~$R_{sun}$, which provides the best fit to the dimming at the end of impulsive phase for 1 October 2011 event. 
The red meridional plane passes through the Sun's center O, the event source region C on the solar surface, and the North Pole. The green plane goes through the point C and is parallel to the Sun’s equatorial plane. The radial direction along line C\textquotesingle~ lies by definition in the red meridional plane, and its projection in the equatorial plane is shown by the dashed orange line. The dashed blue line shows the projection of the cone central axis CP (i.e, the CME direction in 3D) to the meridional plane, and the magenta dashed line shows the corresponding projection to the equatorial plane. As can be seen from Figure~\ref{plane_oct_end}, the CME cone is deflected in the South-East direction. 

By convention, we take the angles in the North/South direction and in the West/East direction as positive/negative in the respective planes. We obtain that the 1 October 2011 CME is inclined from the radial direction by $21^{\circ}$ to the South (i.e, in the meridional plane) and $12^{\circ}$ to the East (in the equatorial plane). This example demonstrates how the CME direction and its inclination from the radial propagation can be estimated early in its evolution using the information from coronal dimmings.

\begin{figure}[!htb] 
	\centering
\includegraphics[width=0.48\textwidth]{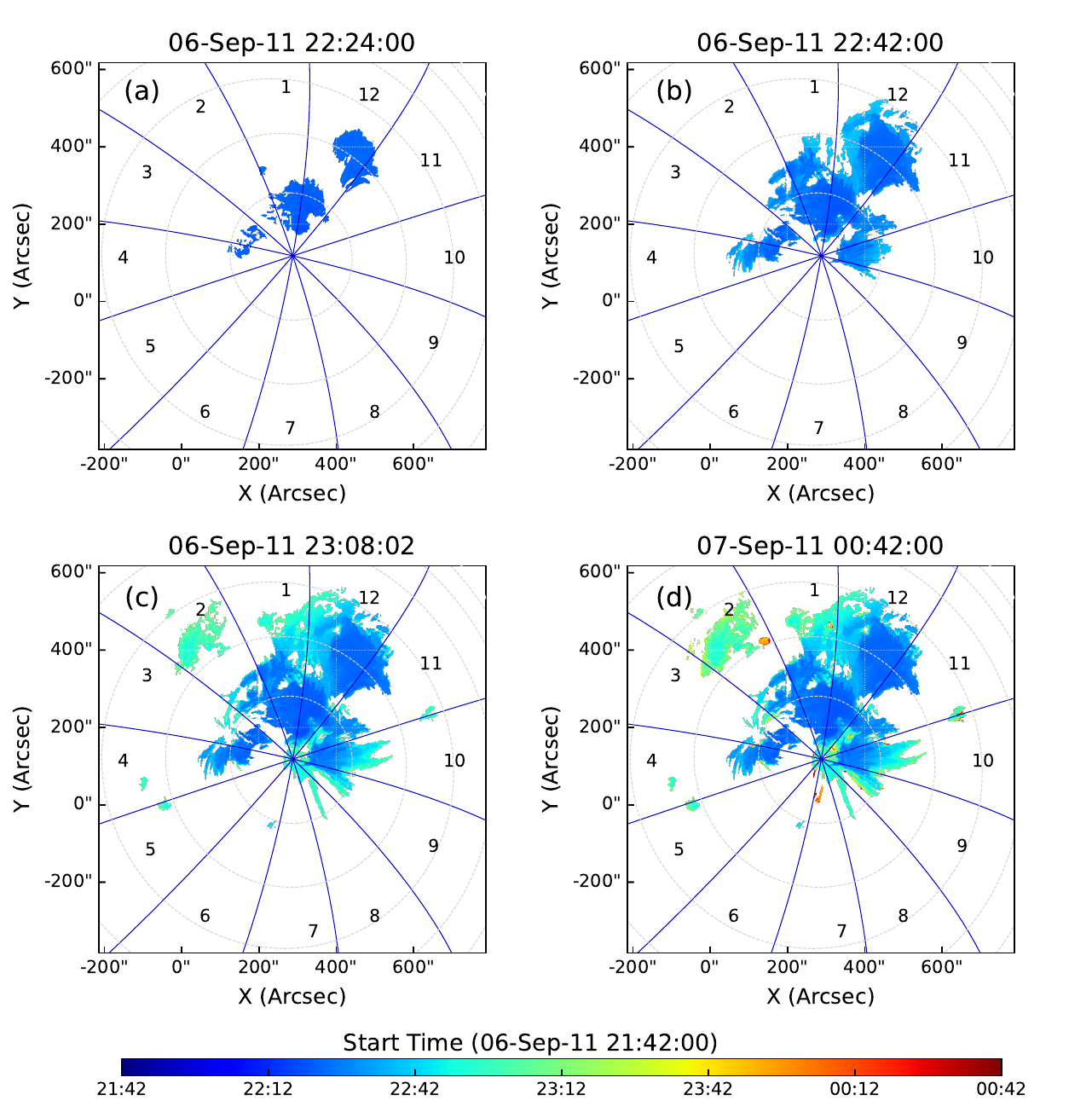}
	\caption{Dimming evolution for the 6 September 2011 event at (a) the maximum of the impulsive phase (12 minutes from the start of the event) (b) 30 minutes from the start of the event, (c) the end of the impulsive phase (44~minutes after the maximum of the impulsive phase), and (d) 2.5 hours from the start. The blue lines show 12 angular sectors and the color bar indicates when each dimming pixel was first detected.} 
	\label{sep6_dimming}
\end{figure}

\begin{figure}[h] 
\begin{subfigure}{\columnwidth}
  \centering
  \includegraphics[width=0.7\linewidth]{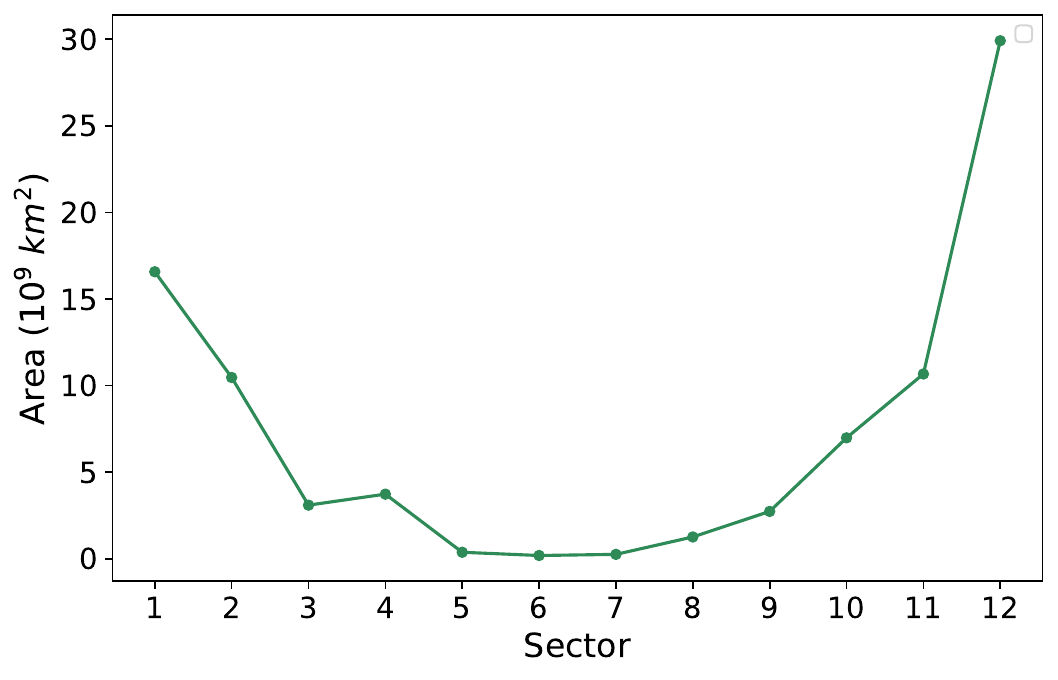}
  \caption{}
  \label{sep6_area_sector_maximum} 
  \includegraphics[width=0.9\linewidth]{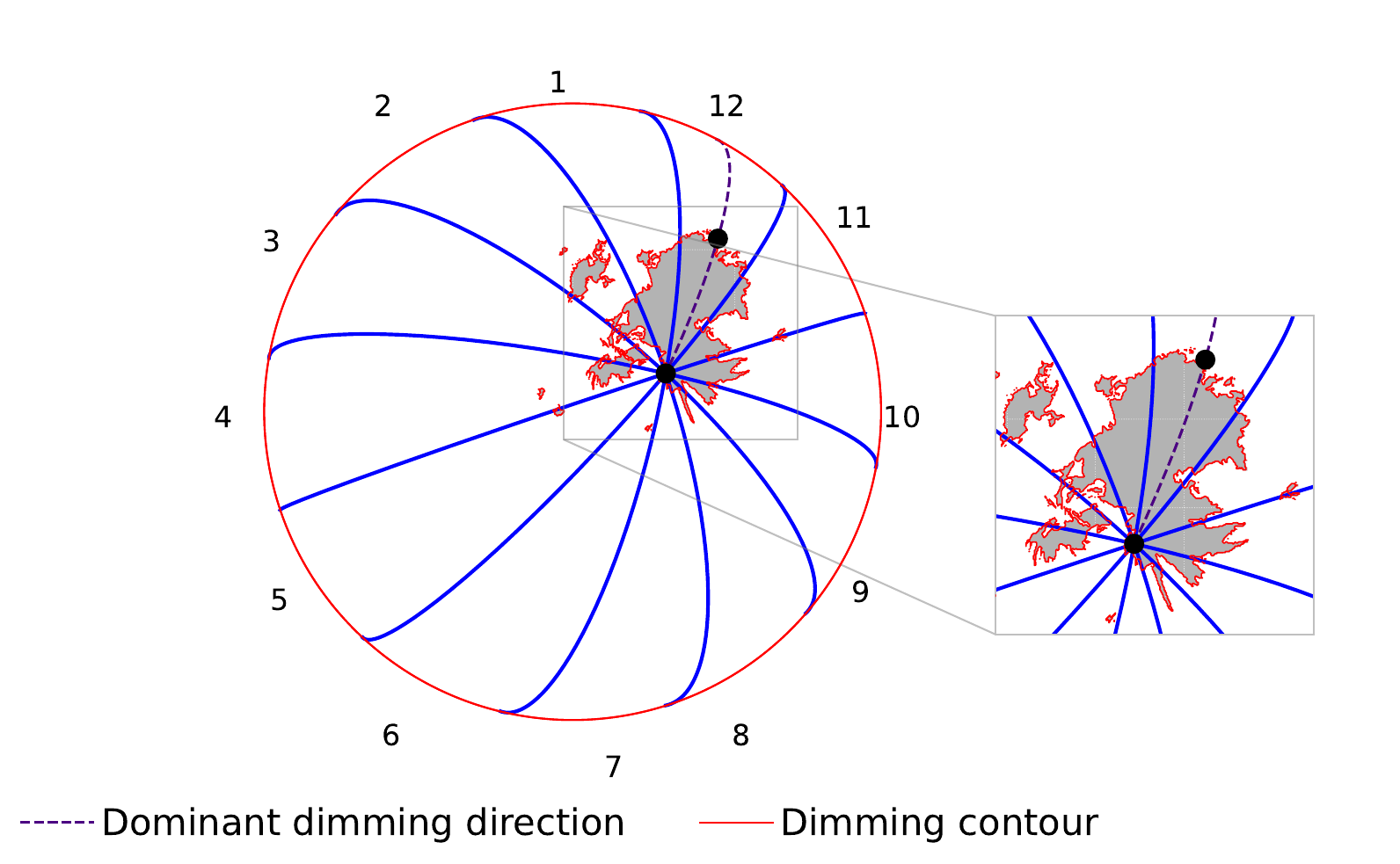}
  \caption{}
  \label{dominant_sector_sep_maximum} 
  \end{subfigure}
\caption{Dominant direction of dimming development for the end of impulsive phase for the 6 September 2011 event. The format is the same as for Figure~\ref{oct_dimming_area_mask}.
The largest dimming area appears in sector 12, i.e. toward North West.}
\label{end_sep_dimming}
\end{figure}

\begin{figure}[!htb]  
	\centering
\includegraphics[width=0.47\textwidth]{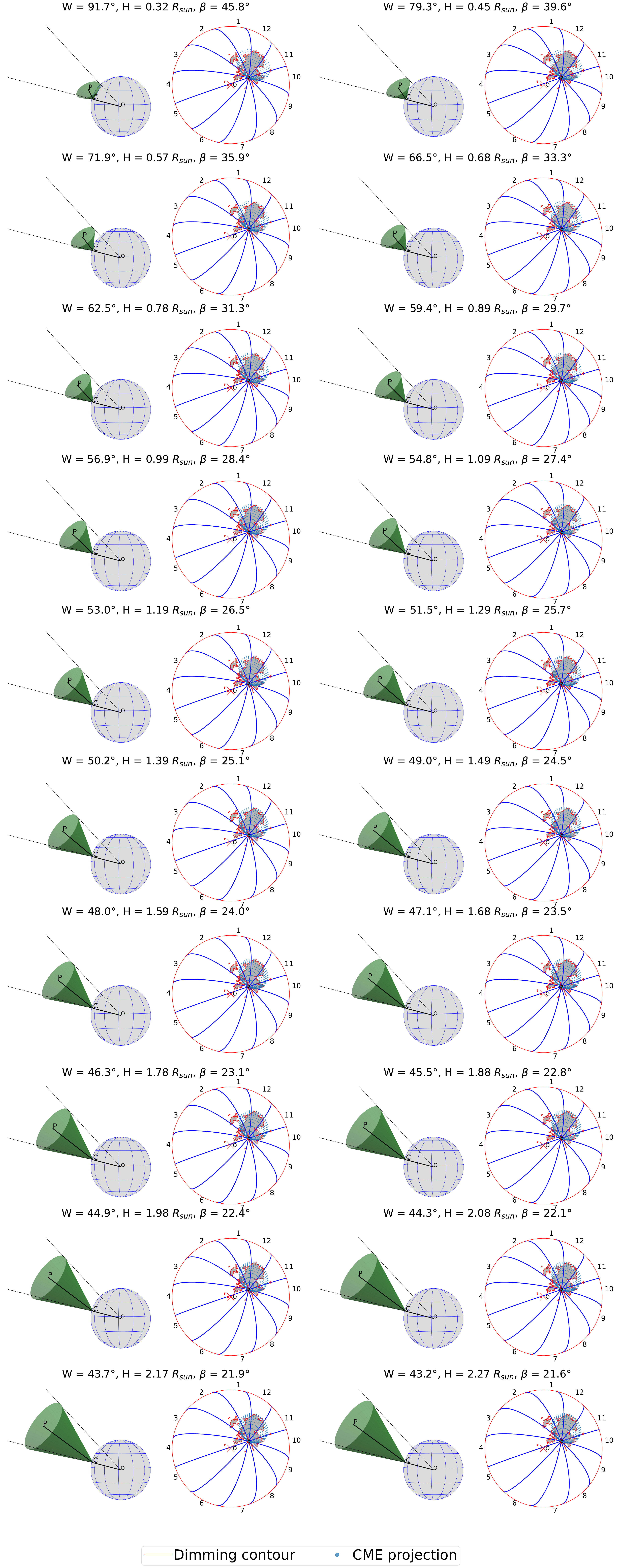}
	\caption{Same as Figure~\ref{oct_proj_end} but for 6 September 2011 at the end of the impulsive phase of the dimming.} 
	\label{sep6_proj_end}
\end{figure}

\section{Results for the 6 September 2011  event}\label{case_study_2}
In this section we apply our DIRECD approach presented in Section~\ref{Methods} to estimate the direction of the early CME propagation from the coronal dimming evolution to the 6 September 2011 event. The authors of \citet{prasad2020magnetohydrodynamic} analyzed this event in detail and reported that the X2.1 class flare occurred close to the disk center at N14$^\circ$ W18$^\circ$, associated with a fast halo CME speed of 990 km/s. The impulsive phase of the flare started at 22:12 UT and reached its
peak around 22:21 UT. 

Figure~\ref{sep6_dimming} shows different phases of the dimming evolution within the first three hours after the start of the at 22:12~UTC. As can be seen, the dimming is mainly 
expanding towards the North-West direction in sectors 1 and 12. 
As for the 1 October 2011 event, we focus our analysis on the dimming extent on the end of the impulsive phase (shown in panel c). The estimates of the dominant direction of dimming development are presented in Figure~\ref{end_sep_dimming}. Panel~(a) shows the resulting dimming area in each sector with the the largest area in sector 12, and panel~(b) gives the cumulative dimming pixel mask at the end of impulsive phase together with the source location and dimming edge in the sector of dominant dimming direction (black dots).

\begin{figure}[h!]
	\centering
\includegraphics[width=0.48\textwidth]{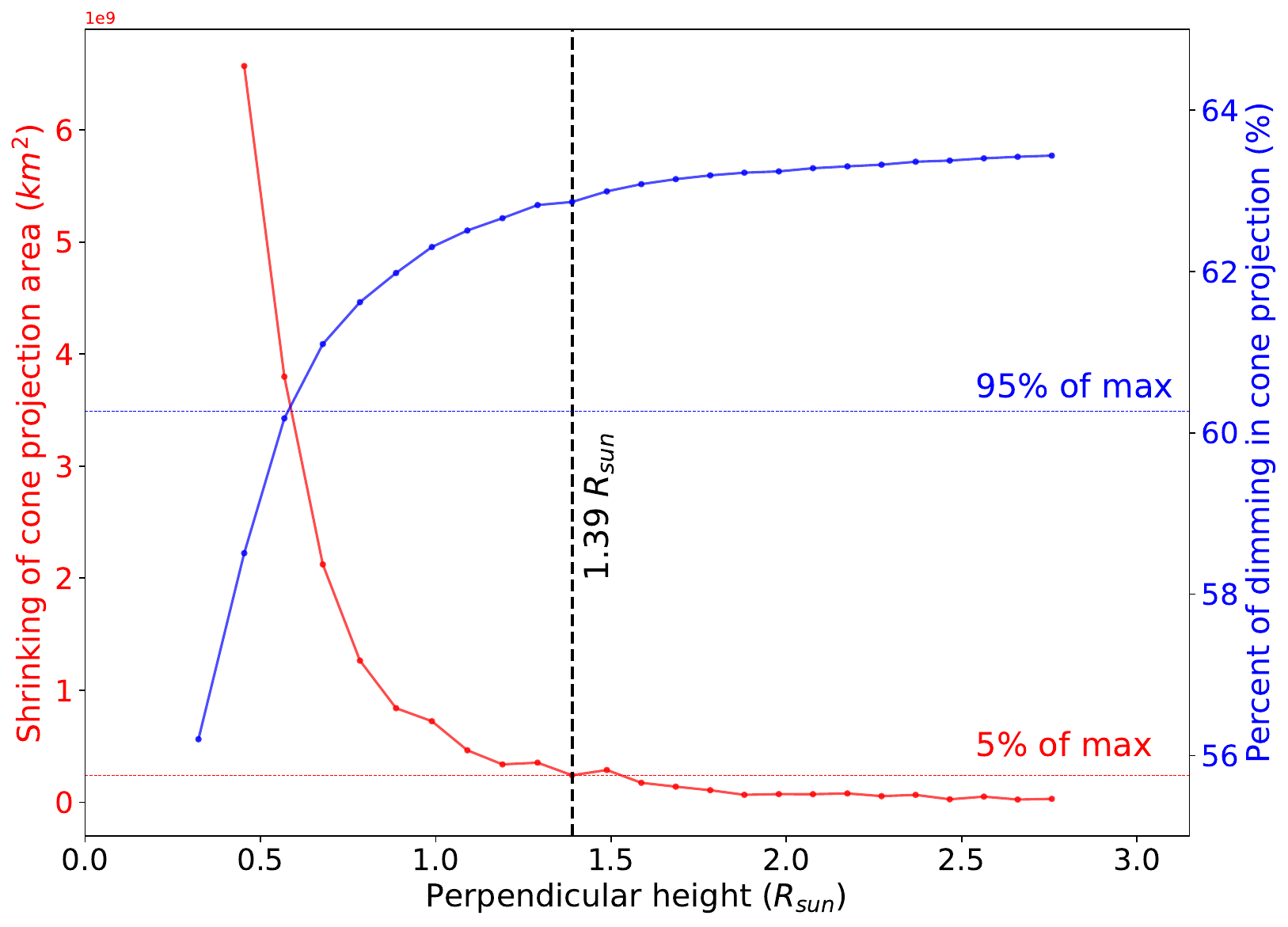}
	\caption{Same as Figure~\ref{dim_area_oct2011} but for 6 September 2011. The best fit is obtained for the cone with a height of 1.39~$R_{sun}$, width of 50.2$^\circ$ and inclination angle of 25.1$^\circ$.} 
	\label{dim_area_sep2011}
\end{figure}

\begin{figure}[h!]
	\centering
\includegraphics[width=0.48\textwidth]{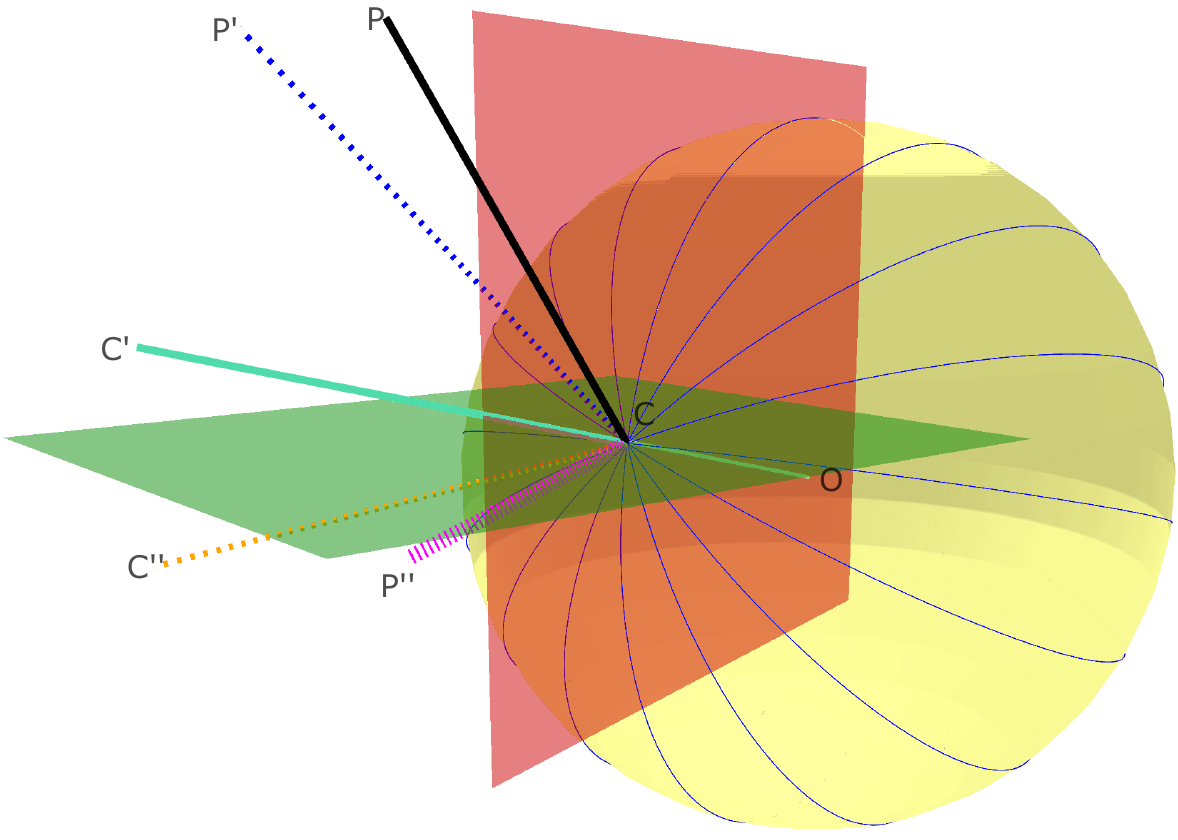}
	\caption{Same as Figure~\ref{plane_oct_end} but for the 6 September 2011 event.} 
	\label{plane_sep_end}
\end{figure}

Figure~\ref{sep6_proj_end} shows twenty generated CME cones 
ranging over heights of 0.32--2.27~$R_{sun}$, widths of 43.2--91.7$^\circ$ and inclination angles of 21.6--45.8$^\circ$ (columns 1 and 3) and their orthogonal projections onto the solar sphere (columns 2 and 4), which are confined within the dimming edges in the sector of dominant dimming direction. To find the best fit cone, we evaluate Figure~\ref{dim_area_sep2011}, which shows the consecutive differences in projection area of generated cones (red, left Y-axis) and percentage of dimming area in projection to the total projection area (blue, right Y-axis) for end of dimming impulsive phase. (This is analogous to Figure \ref{oct_proj_end} for the 1 Oct 2011 event)
For this case, we obtain the best fit parameters for the cone to be: height of 1.39~$R_{sun}$, width of 50.2$^\circ$ and inclination angle of 25.1$^\circ$. 

In Figure \ref{plane_sep_end}, we project the resulting CME direction onto meridional and equatorial planes. We obtain that the 6 September 2011 CME is inclined $22^{\circ}$ North (the meridional plane) and $15^{\circ}$ 
West (the equatorial plane) from the radial direction.

\section{Validation of DIRECD with 3D CME reconstructions}\label{3D_CME}
In this section, we validate the DIRECD method with the CME propagation direction derived from 3D reconstructions of the extended CME bubble for the 1 October 2011 event using STEREO-A(head) and STEREO-B(ehind) EUV images low in corona, where we assume that the CME is still connected to the dimming (Section~\ref{3d_EUV}). Additionally, we analyze and compare the resulting 3D cone from the DIRECD method for the 6 September 2011 event with GCS modeling using STEREO-A/STEREO-B COR 2 white-light images higher up in corona (Section~\ref{3d_GCS}).  

\subsection{3D reconstructions of extended CME loops in the EUV low corona}\label{3d_EUV}
The Solar Terrestrial Relations Observatory \citep[STEREO;][]{Kaiser2008} allows us to perform stereoscopic imaging of the Sun and to compare the obtained stereoscopic CME data with the direction estimates that resulted from DIRECD based on the dimming observations. By taking advantage of the different viewpoint of the STEREO-A and STEREO-B spacecraft, we derive the 3D reconstructions of the extended CME bubble for the 1 October 2011 event using the tie-pointing and triangulation techniques \citep{Inhester2006, Thompson2009, Liewer2009}. We identify identical features in images from the Extreme Ultraviolet Imagers (EUVI; \citep{howard2008sun}) onboard STEREO-A and STEREO-B and employ an algorithm of 3D reconstructions based on epipolar geometry, as described in detail in \citep{Podladchikova2019three}. 

\begin{figure}[h]
\centering	\includegraphics[width=0.5\textwidth]{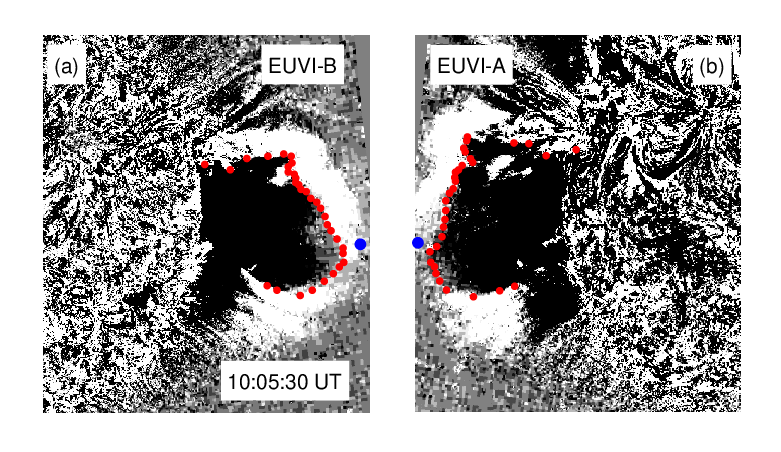}
	\caption{Three-dimensional reconstruction of the extended CME bubbles. Red/blue dots show the inner/outer parts of the CME bubble  matching the same features on both base-difference 195~\AA~ EUVI-B (a) and EUVI-A (b) images.} 
	\label{STEREO_AB}
\end{figure}

\begin{figure}[h]
\begin{subfigure}{\columnwidth}
    \centering
    \includegraphics[width=0.7\textwidth] 
{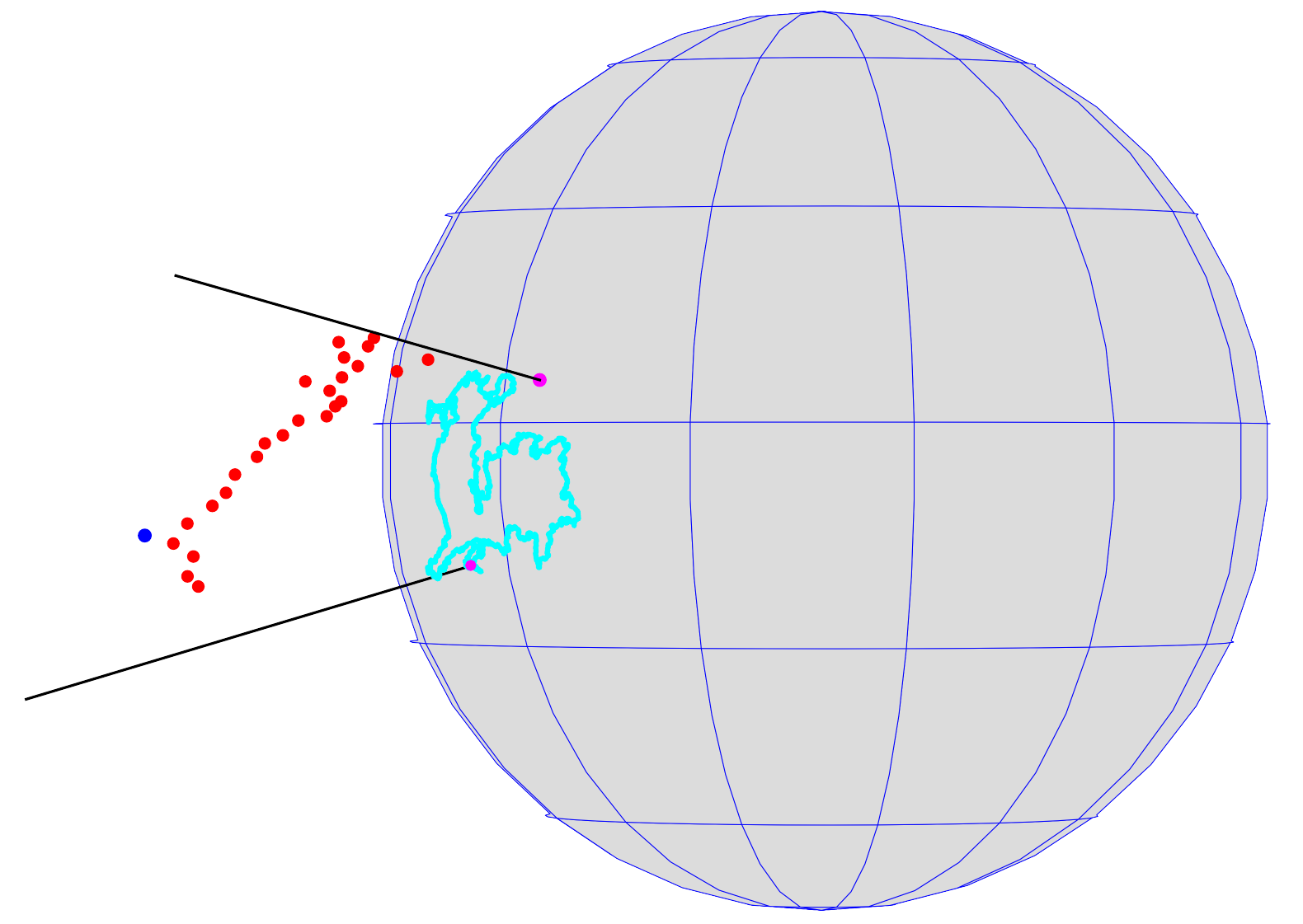}
    \caption{}
    \label{reconstructed_3D}
\end{subfigure}    
\begin{subfigure}{\columnwidth}   
    \centering
    \includegraphics[width=0.6\linewidth]
{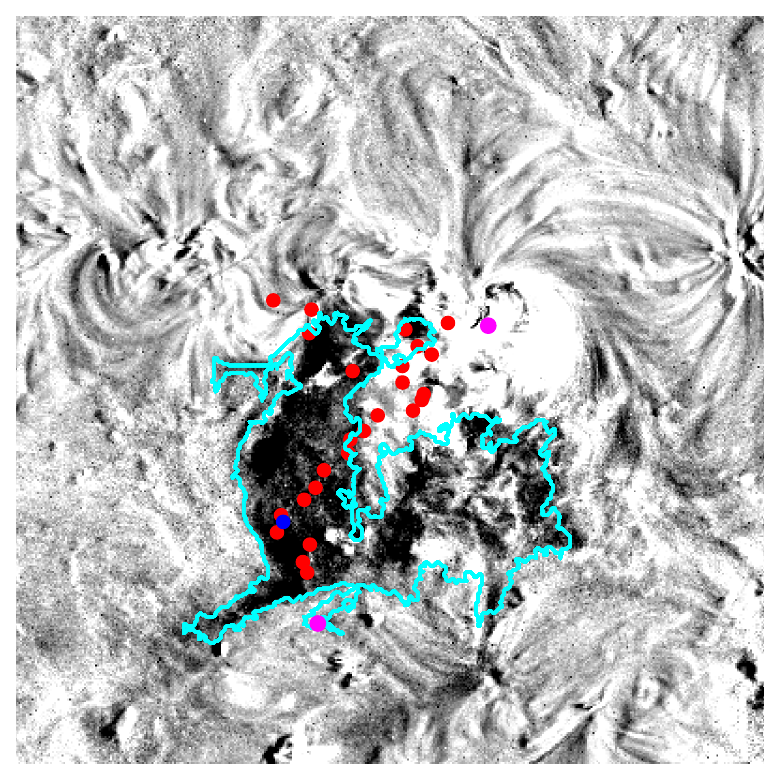}
   \caption{}    \label{dimming_projection_points_oct2011} \end{subfigure}
\caption{Reconstructed CME bubble in 3D (panel~(a), red/blue dots) and its orthogonal projections onto the solar sphere (panel~(b), red/blue dots) together with the boundary of the identified dimming region (panels~(a)~and~(b), cyan). Red/blue dots show the inner/outer parts of the CME bubble and magenta dots indicate the source location and dimming edge along the direction of dominant dimming evolution (panels~(a)~and~(b)). Black lines in panel~(a) have the same meaning as in Figures~\ref{ensemble_cones}~and~\ref{oct_proj_end}. Panel~(b) shows the logarithmic base-ratio SDO/AIA 211~\AA~image at 10:26 UT.}
\label{oct_dimming_3D}
\end{figure}

\begin{figure}[!htb]
\begin{subfigure}{\columnwidth}
    \centering
    \includegraphics[width=0.59\textwidth] 
{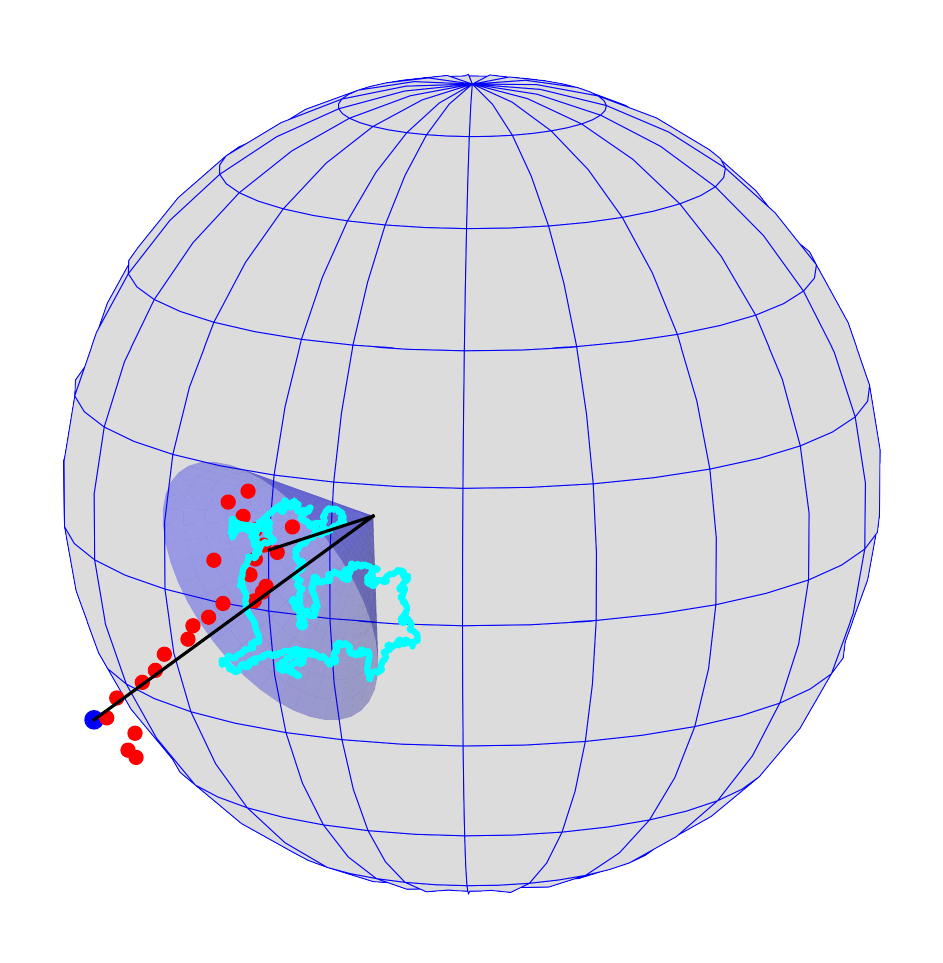}
    \caption{}
    \label{inverse_cone_modelling_oct2011}
\end{subfigure}    
\begin{subfigure}{\columnwidth}    
  \centering
    \includegraphics[width=0.59\linewidth]
 {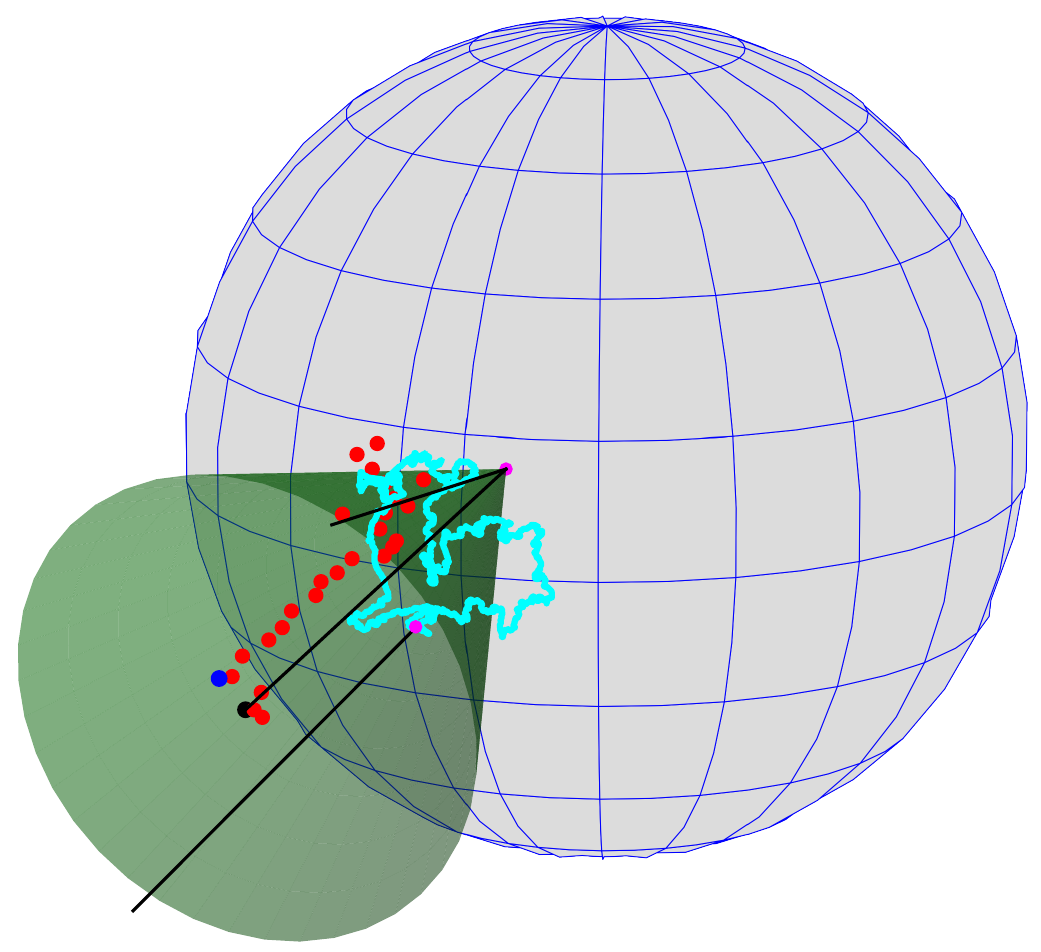}
   \caption{}
    \label{cone_best_fit_oct2011}  
\end{subfigure}
\begin{subfigure}{\columnwidth}   
    \centering
    \includegraphics[width=0.6\linewidth]
{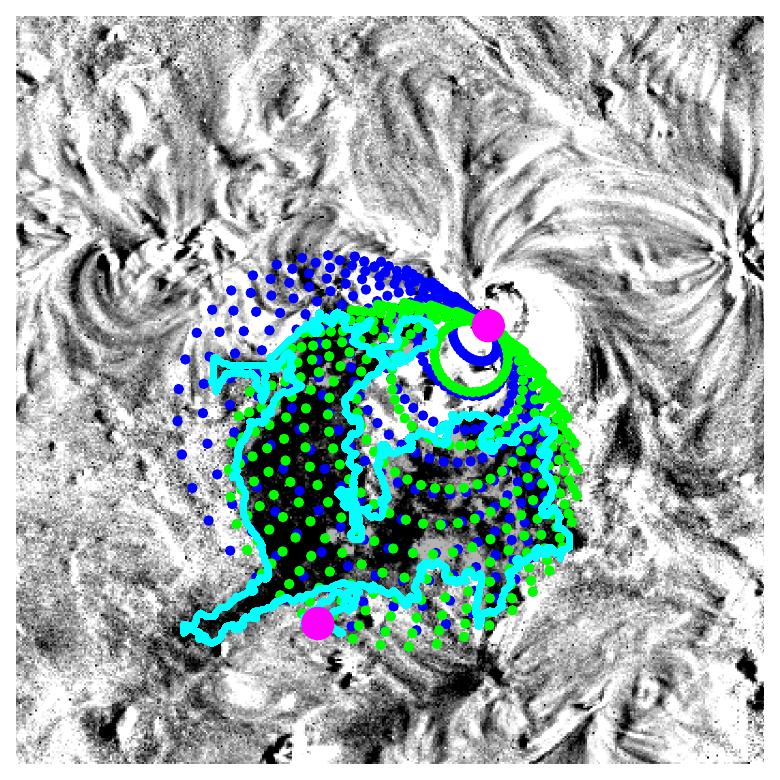}
   \caption{}    \label{dimming_projection_points_oct2011} \end{subfigure}
\caption{Comparison of the CME cones reconstructed from forward modelling and DIRECD. (a) Forward modelled CME cone assuming we know the main direction of the CME propagation from the 3D reconstructions of the CME bubble (b) Best fit CME cone resulted from DIRECD method for 1 October 2011. Red/blue dots in panels (a) and (b) show the reconstructed inner/outer parts of the CME bubble, cyan boundaries indicate the segmented dimming together with the source location and dimming edge along the direction of dominant dimming evolution (magenta dots). Panel (c) shows the logarithmic base-ratio SDO/AIA 211~\AA~image together with the orthogonal projections of the CME cone from the forward modelling (blue dots) and that from our method DIRECD (green dots) together with the segmented dimming (cyan).}
\label{oct_dimming_forward_modelling}
\end{figure}

Figure~\ref{STEREO_AB} shows the expanding CME bubble at 10:05~UT as viewed with base-difference 195~\AA~ EUVI images of STEREO-B (a) and STEREO-A (b). As we presume the CME is connected to the dimming at a certain height, we perform stereoscopic estimates for the inner part of the CME bubble on the border between the end of the visible off-limb dimming and the bright CME outer edge and show the matching features on EUVI-B and EUVI-A with red dots. Additionally, we derive the stereoscopic reconstructions of the outer edge of the CME bubble, indicating its main propagation direction by the blue dot in both EUVI-B and EUVI-A images. 

Figure~\ref{oct_dimming_3D}a shows the resulting 3D reconstructions, where red and blue dots show the inner and outer parts of the CME bubble, with the height reaching 550~Mm. Cyan boundaries show the segmented dimming, magenta dots indicate the source location and dimming edge along the direction of the dominant dimming evolution, through which we construct the edge black lines extended into 3D. As can be seen from Figure~\ref{oct_dimming_3D}a, the reconstructed inner part of the CME bubble (red points) together with the outer edge (blue point) is located right above the dimming. Figure~\ref{oct_dimming_3D}b shows the orthogonal projections of the reconstructed points of the CME bubble (red and blue dots) together with the segmented dimming (cyan) on top of a logarithmic base-ratio SDO/AIA 211~\AA~image. As can be seen from Figure~\ref{oct_dimming_3D}b, the projections are mainly concentrated in the center of the dimming, which justifies our inital assumption to link the 2D dimming with the 3D CME bubble. Projection points in between the dimming edge and source region are probably related to the part of non-detected dimming areas covered by the solar flare at certain time steps during the evolution. As can be also seen from the Figure~\ref{oct_dimming_3D}b, the projection points of the CME bubbles are mainly located not in the center of the dimming, but in its left (Eastern) part, which is presumably explained by reconstructing the non-central slice of the CME bubble. 

The 3D reconstruction of the CME bubble allows us to perform forward modelling and reconstruction of the CME cone and its orthogonal projections, assuming we know the main direction of CME propagation, which goes along the line connecting the source region and outer edge of the CME bubble (Figure~\ref{oct_dimming_forward_modelling}a).The half width of the cone (45$^\circ$) in this case is determined by the angle between the line connecting the Sun's center and the source location, and the line connecting the Sun's center to the assumed main direction of the CME propagation (which is represented by the reconstructed outer edge of the CME bubble indicated by the blue dot). The width is thus 90$^\circ$. For comparison, in Figure~\ref{oct_dimming_forward_modelling}b, we show the CME best fit cone resulting from our DIRECD method (see Section~\ref{Dimming_CME_direction}) together with the 3D reconstructions of the CME bubble. As can be seen from Figure \ref{dimming_projection_points_oct2011}, the central axis of the cone reconstructed with DIRECD and the outer edge of CME bubble (blue point) are located in proximity to each other, though there is some shift of the CME outer edge representing the non-central slice of the CME bubble. This is also seen in Figure~\ref{oct_dimming_forward_modelling}c, which shows the orthogonal projections of the CME cone from the forward modelling (blue dots) and that derived using DIRECD (green dots) together with the segmented dimming (cyan) plotted onto a logarithmic base-ratio SDO 211~\AA~image. Note that for the forward modelled cone, the projections become wider with increase in height in contrast to the cone reconstructed by DIRECD that becomes narrower with increase in height. While blue dots (orthogonal projections from the forward modelling) are shifted, green dots (orthogonal projections from the best fit cone) are centered around the dimming. This is an additional argument of the reconstructed outer edge of CME bubble to represent a part of non-central slice of CME bubble, which is in general going to the same direction as estimated with our approach. 

\subsection{3D reconstructions of the white-light CME in the higher corona with GCS}\label{3d_GCS}
To derive the 3D structure and direction of the CME bubble for the 6 September 2011 event, we use GCS reconstructions available from the HELCAT Kincat catalog \footnote{\url{https://www.helcats-fp7.eu/catalogues/wp3_kincat.html}} using STEREO-A/STEREO-B COR 2 white-light data. The parameters for the GCS reconstruction are as follows: heliocentric Longitude = 40$^\circ$, heliocentric Latitude  = 34$^\circ$, tilt = 27$^\circ$, aspect ratio $r$ = 0.42, half-angle measured between the apex and the central axis of a leg of the croissant $\alpha/2$ = 35.5$^\circ$, and the height of the croissant at the end of the dimming impulsive phase at 23:08~UT = 6~$R_{sun}$. 

\begin{figure}[h!]
	\centering
\includegraphics[width=0.48\textwidth]{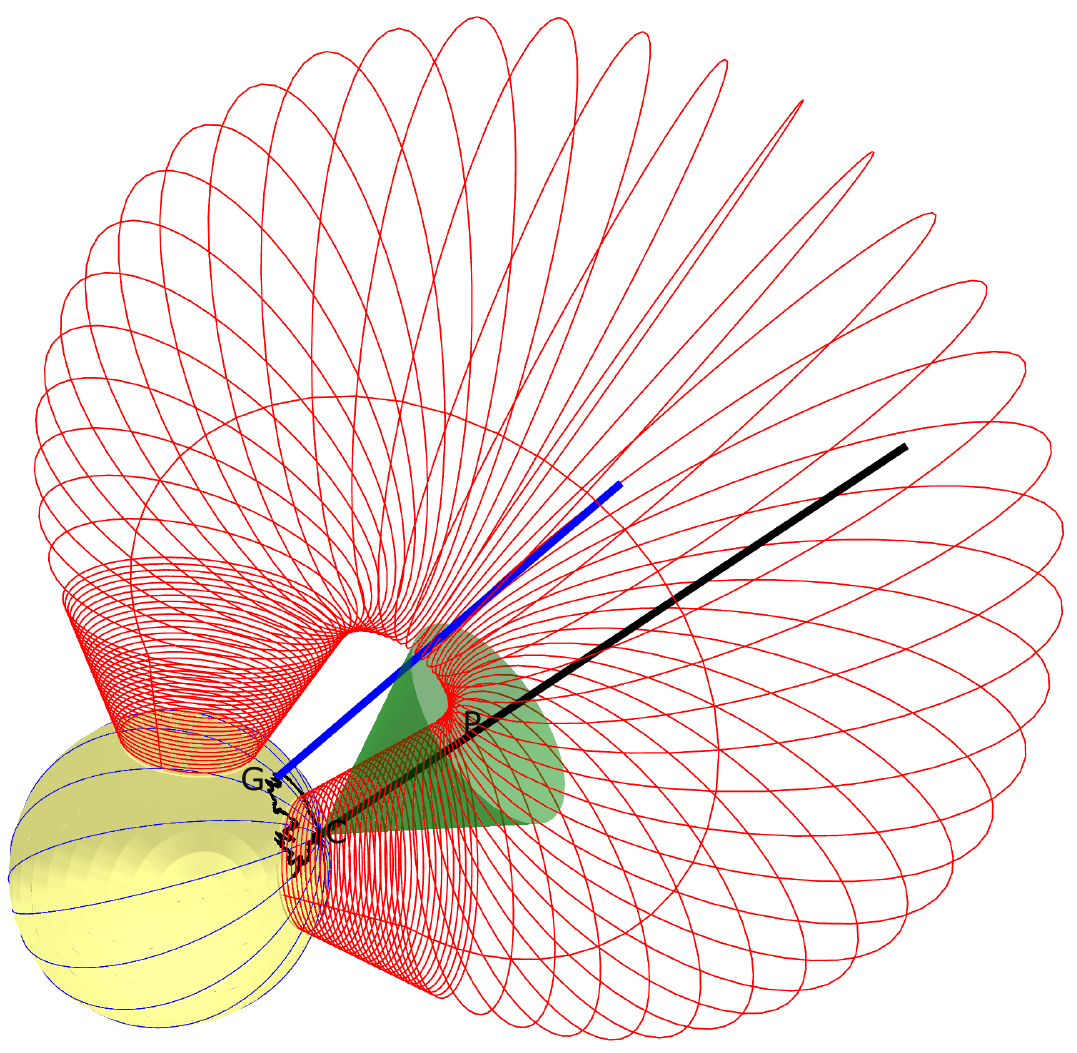}
	\caption{Comparison of GCS and DIRECD method for 6 September 2011 at the end of the dimming impulsive phase (23:08~UT). The red mesh shows the reconstructed GCS croissant, the blue line is the central axis of GCS. The black line is the central axis of the cone obtained from DIRECD. Points C, G are flare and GCS source, respectively.} 
	\label{GCS_Sep2011}
\end{figure}

With the Python code provided in \cite{johan_l_freiherr_von_forstner_2021_5084818}, we plotted the resulting GCS croissant (red mesh) together with the 3D cone (green) obtained with the DIRECD method shown in Figure~ \ref{GCS_Sep2011}. The blue line indicates the central axis of the GCS reconstruction, while the black line is the central axis of the 3D cone from DIRECD. Points C and G mark the flare and GCS source, respectively. Since GCS assumes radial propagation of the CME from the GCS source and also may include uncertainties of the reconstruction, the GCS source G is shifted from the flare source C. We find that the angle between the vector from the Sun center to point G and that from the Sun center to point C is 28 $^\circ$, which allows us to roughly estimate  the inclination from the radial direction (if the CME was propagating radially then the point G should be close or coincide with the point C). This result is close to the inclination angle obtained from the DIRECD method (25$^\circ$) at the end of the dimming impulsive phase. 
Moreover, the angle between the GCS central axis (blue line) and the DIRECD cone central axis (black line) is 5.8 $^\circ$ indicating propagation roughly in the same direction.

In addition, we note that \citet{Chikunova2023} studied the 28 October 2021 event and reported on a good alignment of the dimming contours with the inner part of the GCS croissant located between the two footpoints, as both flare and GCS sources were closely located, indicating a radial (or close to radial) CME propagation direction. In our case, we obtain that the height of the DIRECD CME cone (1.39 $R_{sun}$) is close to the height of the inner part of GCS croissant (1.45 $R_{sun}$). As both the DIRECD CME cone and the GCS croissant are reconstructed at the same time at the end of the dimming impulsive phase (the CME bubble is already seen in coronagraphs at this time), this supports the argument that the CME propagation is connected to the dimming and leaves footprints in the low corona only up to a limited height. These resemblances imply that the best-fit cone from the DIRECD method can be used not only to link the 2D dimming with the 3D CME bubble to provide an estimate of the early CME direction, but also to provide further information for improving estimations of GCS parameters.

\section{Discussion and Conclusions}~\label{Summary}
In this study, we developed a novel method called DIRECD to characterize the early CME propagation direction from the expansion of coronal dimmings. The method assumes a relationship between CMEs and dimmings in the low corona, and the proposed approach consists of the following main steps. First, to understand the overall 3D projection of CME cones, we simulated CMEs in 3D using a geometric cone model and demonstrated that the larger the CME height and width, the larger the area of CME orthogonal projection onto the solar surface, and its shape varies from circles to ellipses depending on the inclination angle from the radial direction and the source location. Second, we estimated the dominant direction of the dimming extent from the evolution of the dimming area. Third, using the derived dominant sector of the dimming evolution on the solar sphere, we solved an inverse problem to reconstruct an ensemble of CME cones at different heights, widths and different inclinations from the radial propagation. Finally, we chose the CME parameters,
for which the CME orthogonal projections onto the
solar sphere best match the dimming geometry and main direction at the
end of its impulsive phase and derived the CME geometry in 3D. 
We applied the proposed approach for two well-observed CME/dimming events of 1 October 2011 and 6 September 2011.

We find that for the 1 October 2011 event the best fit of the CME projections to the dimming extent is a CME cone with height of 1.04~$R_{sun}$ (a height at which the CME still remains connected to the dimming), CME width of 48.7$^\circ$ and inclination angle of 24.3$^\circ$. By projecting the obtained 3D direction onto the meridional and equatorial planes, we obtained that the CME was directed towards the South-East ($21^{\circ}$ South and $12^{\circ}$ East) away from the radial direction. This outcome is in agreement with the study by \citet{temmer2017flare}, where the authors used GCS reconstruction from multi viewpoint coronagraph observations of the CME, and concluded that the associated two-step filament eruption and the CME propagated towards the South-East. They report a near-Earth ICME associated with the event that was detected on 5 October 2011. 

For the 6 September 2011 event, we derived that the best fit of CME projections to the dimming extent results in a CME cone with height of 1.39~$R_{sun}$, width of 50.2$^\circ$ and inclination angle of 25.1$^\circ$.  We find that the CME was inclined towards North-West from the radial direction ($22^{\circ}$ North and $15^{\circ}$ West). This outcome is in agreement with \citet{kay2017using}, who studied this event using 3D reconstructions from STEREO images and found a northward latitudinal deflection and a small westward longitudinal deflection. This event was also associated with an ICME that occured on 9 September 2011. 

We showed that the CME direction from the DIRECD method is in line with that derived from  the 3D tie-pointing of the CME bubble observed in EUV low in the corona and from the GCS 3D modelling of the white-light CME higher up in corona. Thus, our study presents a proof of concept that coronal dimmings can be used as a proxy of the early CME propagation direction. The main advantages of the DIRECD method are a) that it can be used to give an early estimate of CME directions before it may even appear in the coronagraph field-of-view, b) that is does not need multi-viewpoint observations or coronagraphs, and c) that it is particular useful for Earth-directed events which are difficult to assess from coronagraphs based in the Sun-Earth line, as they mostly observe the CME expansion but not its propagation. Moreover, this method may compensate for the lack of STEREO/COR1 observations, which would be helpful for the tracking of small CMEs in the early propagation in the lower corona.

On the other hand, the method has of course also its own limitations.  
We note that this method is only suitable for CMEs
that are initiated below 0.4 Rsun and for their propagation up to
3 Rsun above the surface since higher up no dimming footprint is
detected anymore. These limitations come from the findings from \cite{robbrecht2009no} which found stealth CMEs to occur above 0.4 $R_{sun}$ and \cite{dissauer2019statistics} which found the initiation height of dimming-associated CMEs to be 0.16 $\pm$ 0.13 $R_{sun}$ and the majority of the dimming evolution within 3 solar radii of the associated CME evolution. Thus, the method can not account for any effects that may influence the CME direction further out and in interplanetary space, which is in contrast to the results of GCS reconstructions that can be applied up to distances of 15-20~$R_{sun}$. Further steps in estimating the CME direction from coronal dimmings need to include a larger statistical sample of CME events, which would allow investigating the accuracy and limitations of the method in detail, as well as a study of how often the early CME direction/deflection estimates using the presented method reflect the CME's behaviour further out in the heliosphere.

 \begin{acknowledgements}
S.J., T.P. and G.C. acknowledge support by the Russian Science Foundation under the project 23-22-00242. The authors thank the International Space Science Institute (ISSI) Team on  “Coronal Dimmings and Their Relevance to the Physics of Solar and Stellar Coronal Mass Ejections”  for productive discussions, which helped to advance this study. SDO data is courtesy of NASA/SDO and the AIA, and HMI science teams. The STEREO/SECCHI data are produced by an international consortium of the Naval Research Laboratory (USA), Lockheed Martin Solar and Astrophysics Lab (USA), NASA Goddard Space Flight Center (USA), Rutherford Appleton Laboratory (UK), University of Birmingham (UK), Max-Planck- Institut f\"ur Sonnenforschung (Germany), Centre Spatiale de Li\'ege (Belgium), Institut d'Optique Th\'eorique et Appliqu\'ee (France), and Institut d' Astrophysique Spatiale (France).We thank the referee for valuable comments on this study.
 \end{acknowledgements}

 \bibliographystyle{aa}
\bibliography{My_References.bib}

\begin{appendix}
\section{Simulations with a CME cone model for different source region locations} \label{Appendix_A}
\begin{figure}[h]  
	\centering
\includegraphics[width=0.5\textwidth]{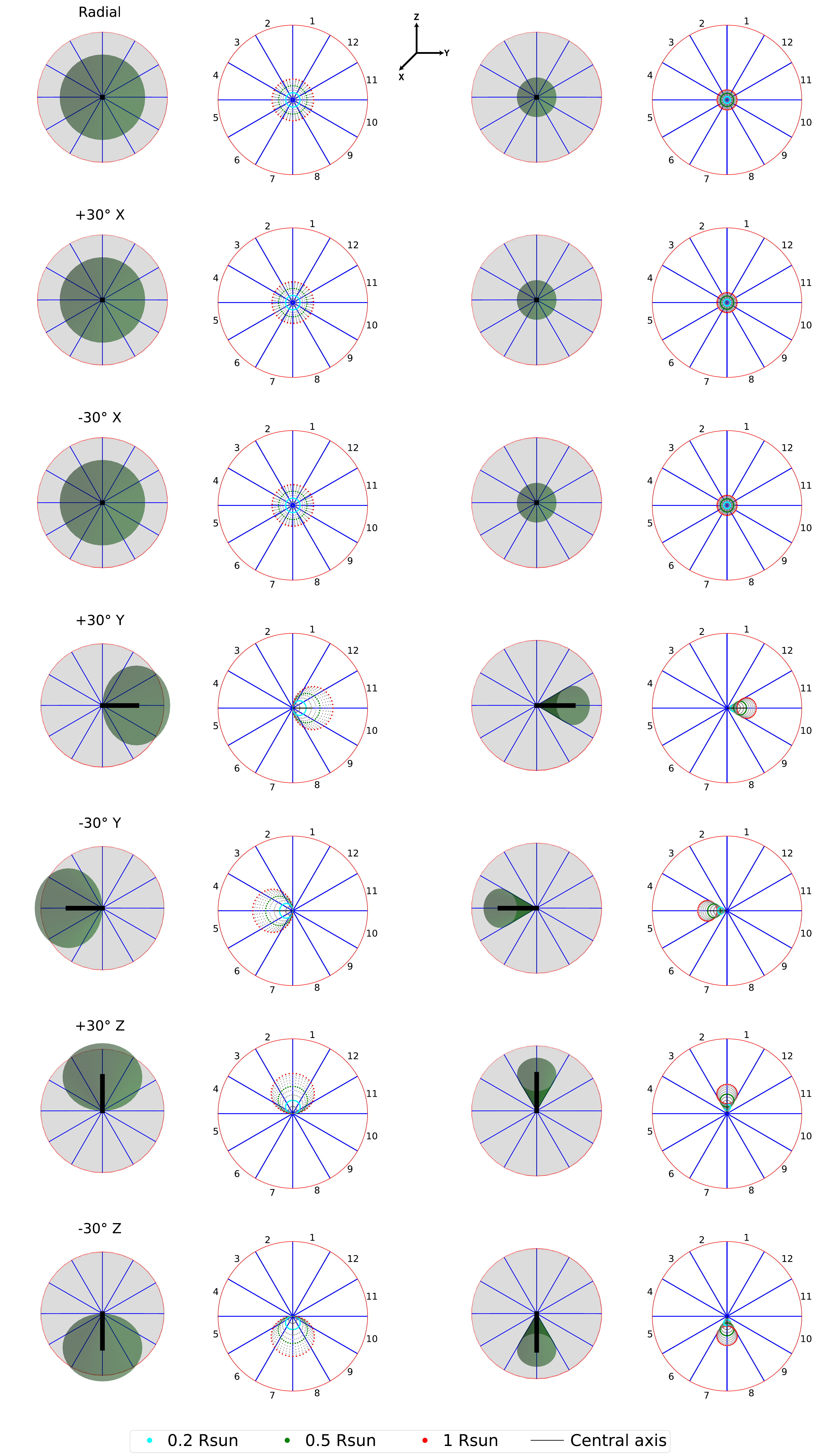}
	\caption{Same as Figure~\ref{sim_1_l}, but for the source region in the Sun center.} 
	\label{sim_cen}
\end{figure}

\begin{figure}[h]  
	\centering
        \vspace{1.6cm}
\includegraphics[width=0.5\textwidth]{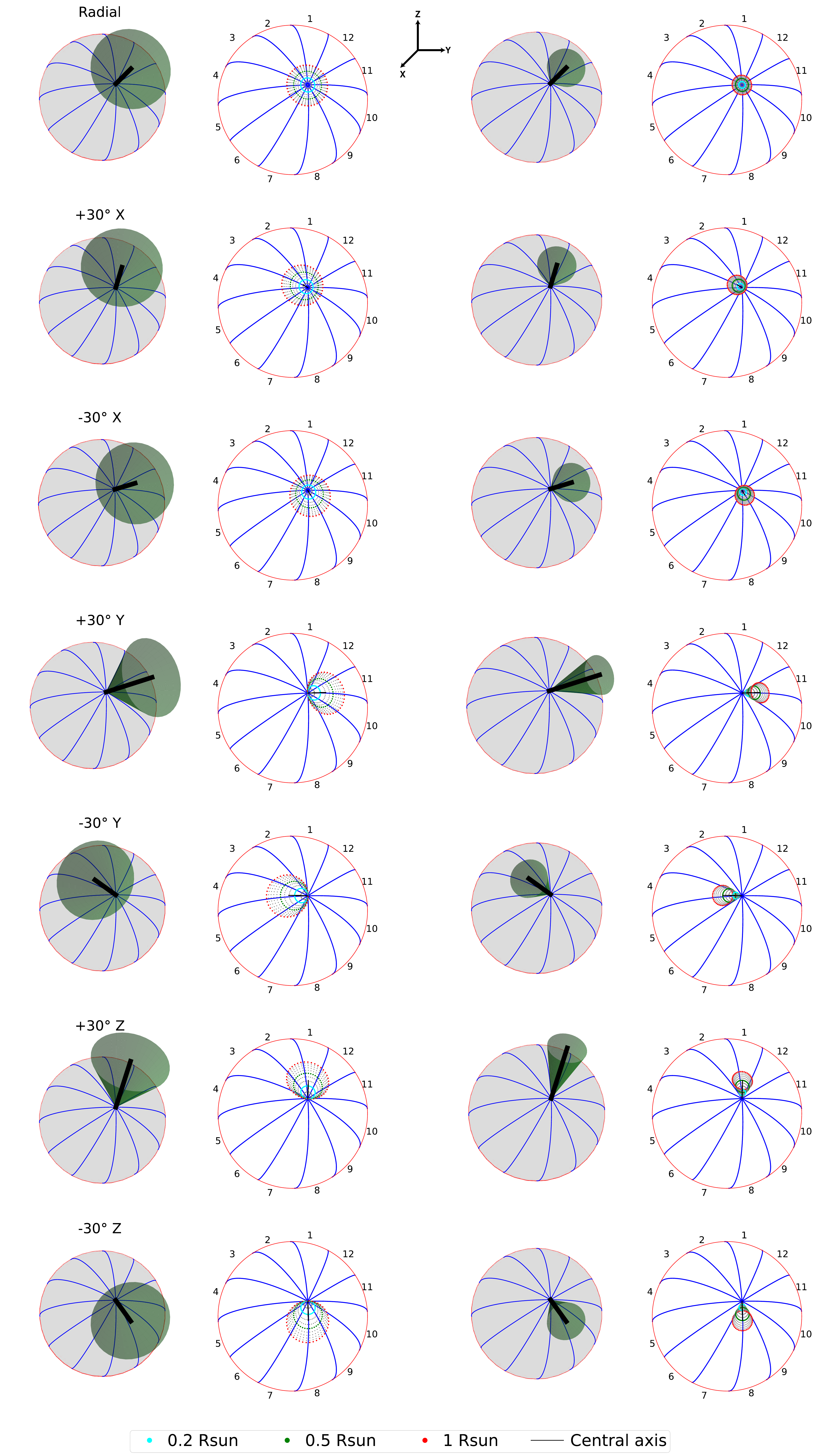}
	\caption{Same as Figure~\ref{sim_3_nl}, but for quadrant 1 (near center)} 
	\label{sim_n1_l}
\end{figure}

\begin{figure}[h!]
	\centering	
	\includegraphics[width=0.5\textwidth]{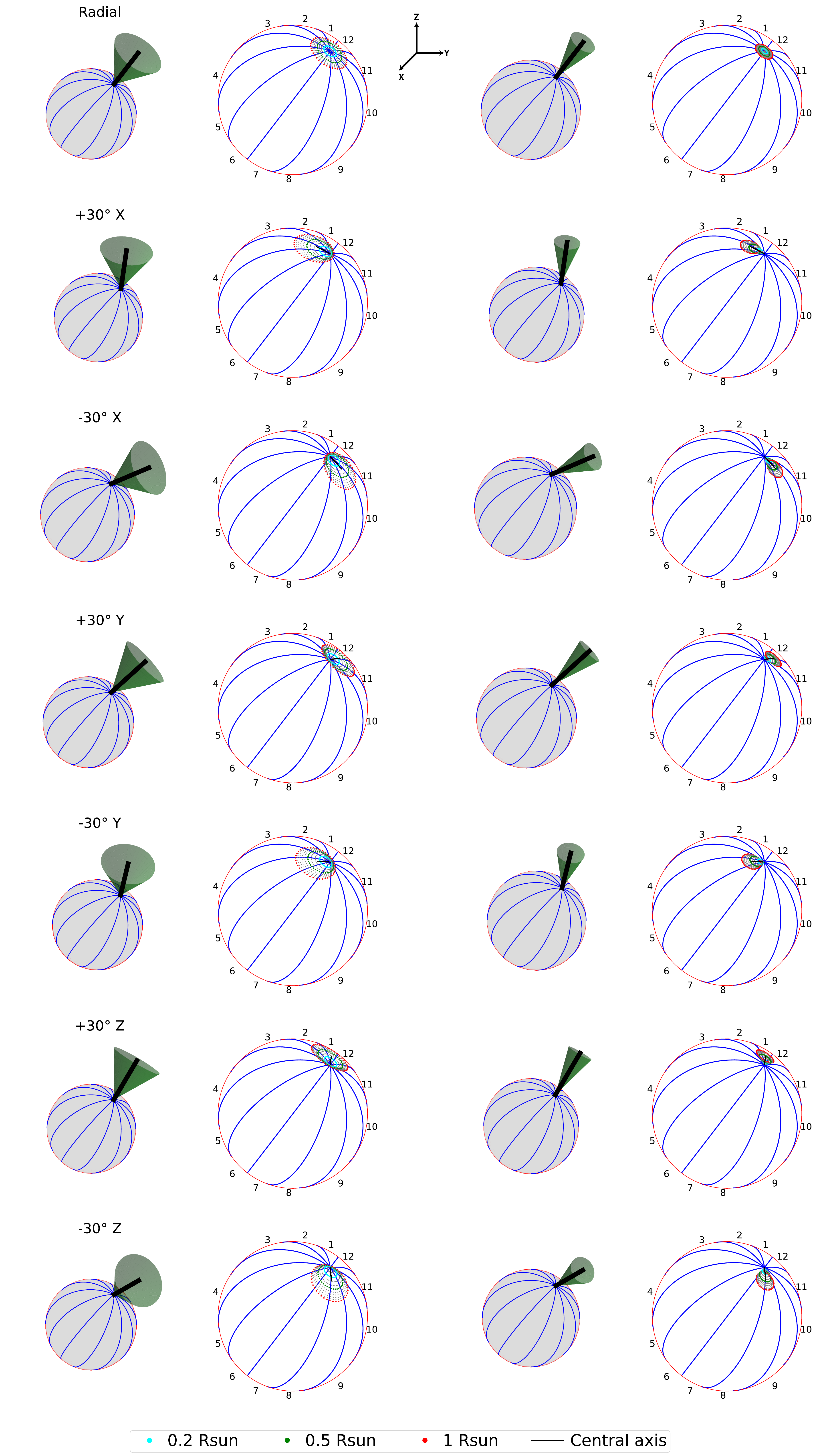}
	\caption{Same as Figure~\ref{sim_3_nl}, but for quadrant 1 (near limb).} 
	\label{sim_1_l}
\end{figure}

\begin{figure}  
	\centering
	
	\includegraphics[width=0.5\textwidth]{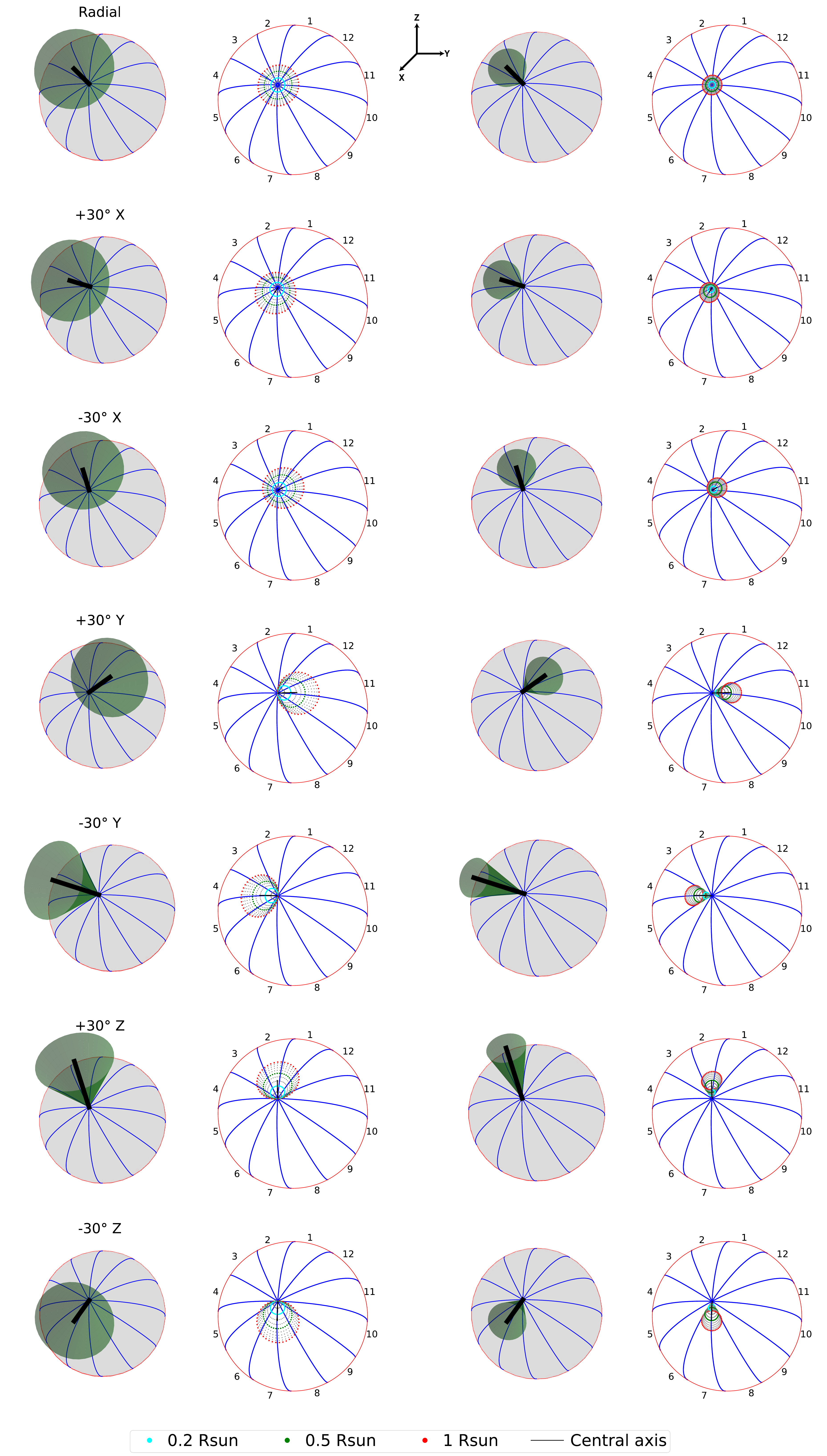}
	\caption{Same as Figure~\ref{sim_3_nl}, but for quadrant 2 (near center).} 
	\label{sim_2_l}
\end{figure}

\begin{figure}  
	\centering
	\includegraphics[width=0.5\textwidth]{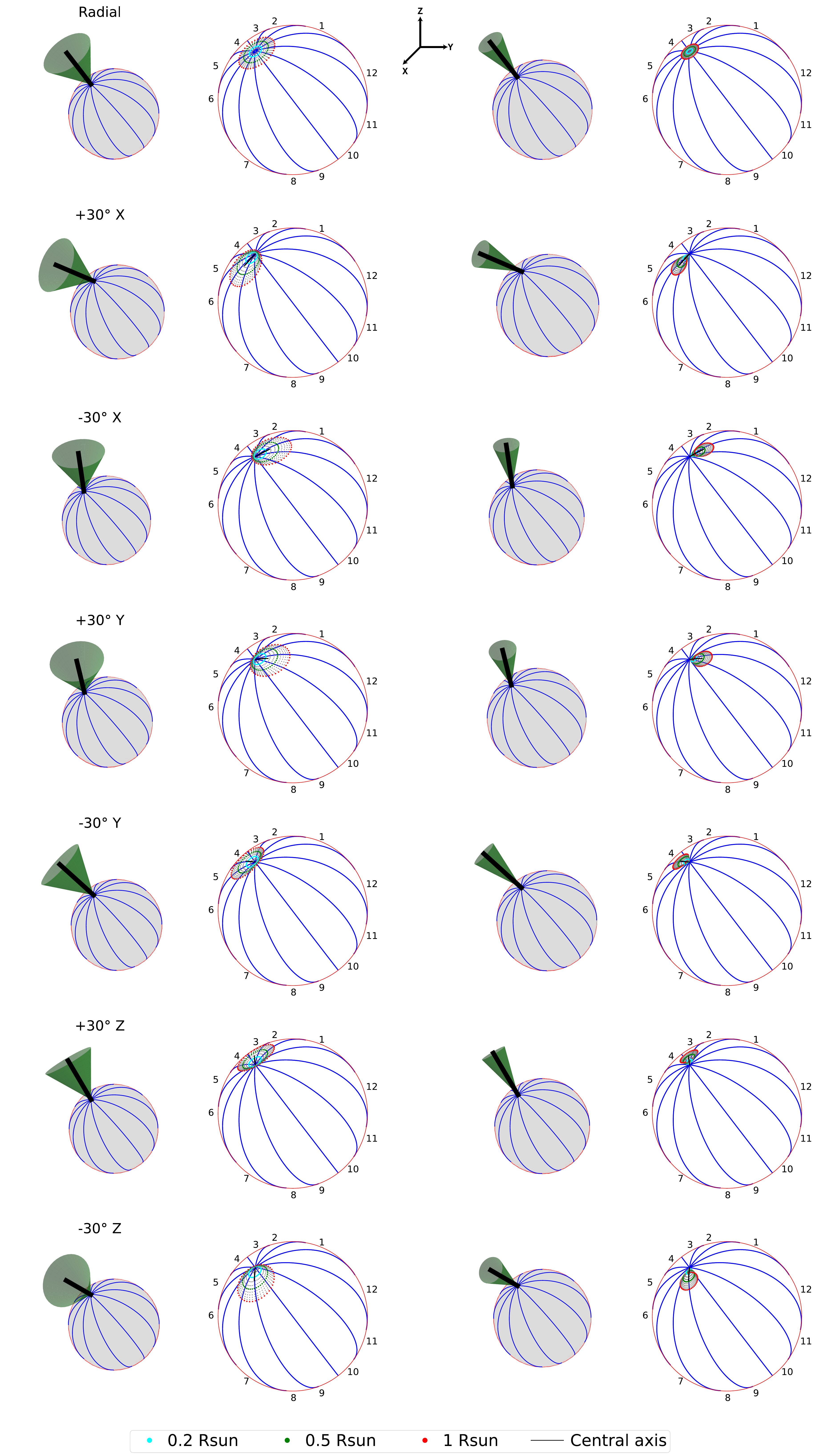}
	\caption{Same as Figure~\ref{sim_3_nl}, but for quadrant 2 (near limb).}
	\label{sim_2_nl}
\end{figure}

\begin{figure}  
	\centering
	\includegraphics[width=0.5\textwidth]{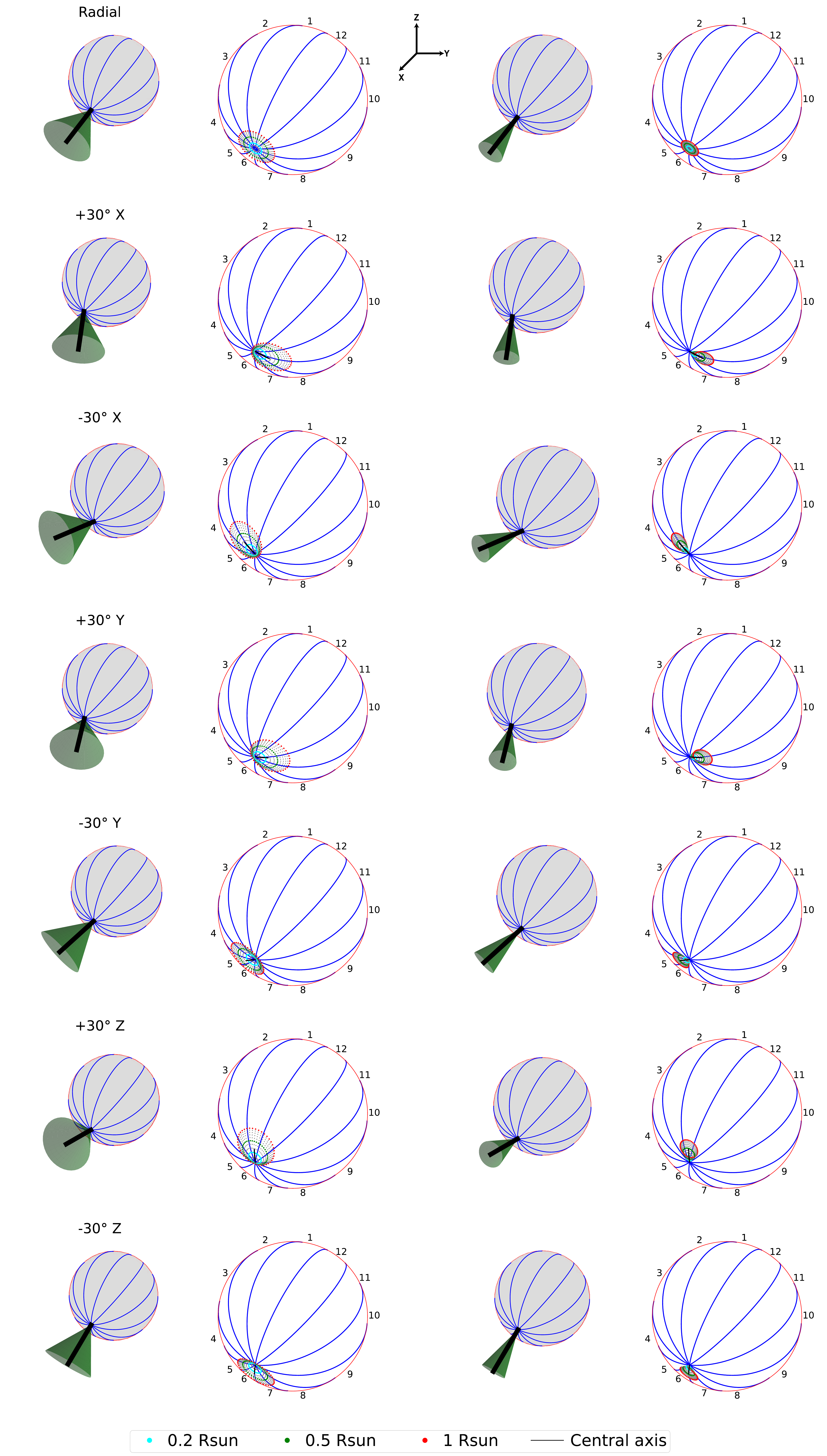}
	\caption{Same as Figure~\ref{sim_3_nl}, but for quadrant 3 (near limb).}  \label{sim_3_l}
\end{figure}

\begin{figure}  
	\centering
	\includegraphics[width=0.5\textwidth]{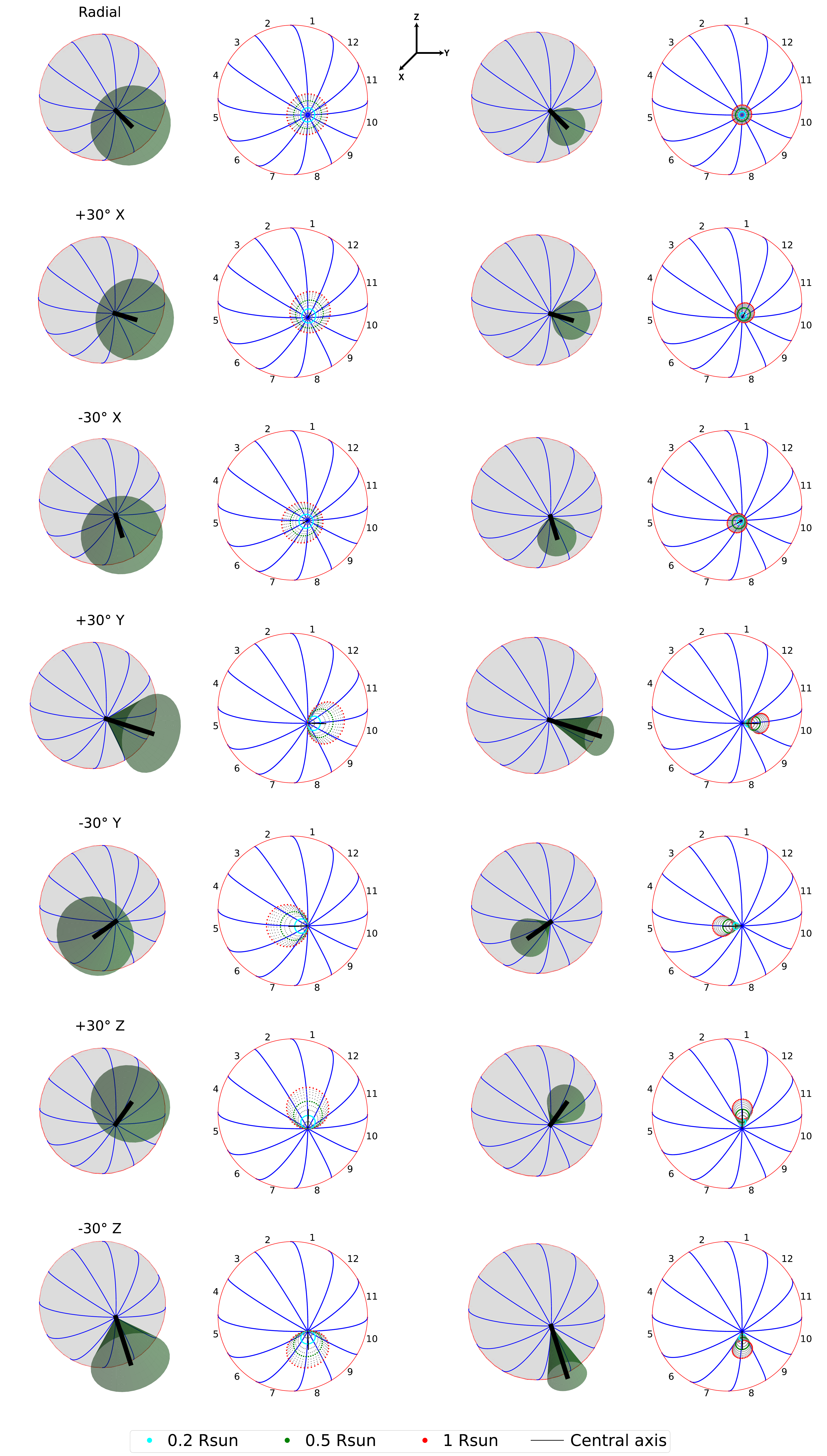}
	\caption{Same as Figure~\ref{sim_3_nl}, but for quadrant 4 (near center).}
	\label{sim_4_nl}
\end{figure}

\begin{figure}  
	\centering
	\includegraphics[width=0.5\textwidth]{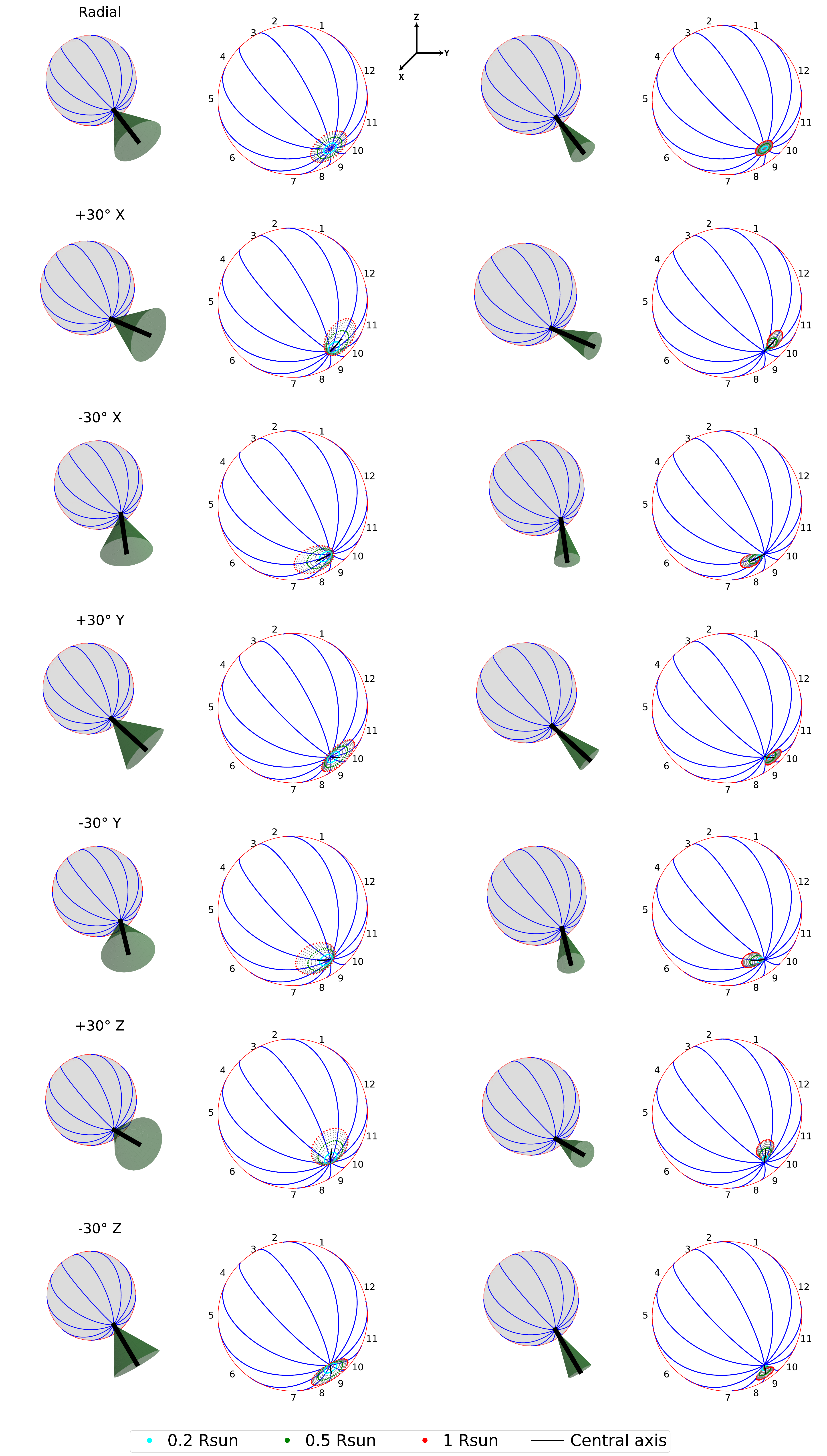}
	\caption{
                 Same as Figure~\ref{sim_3_nl}, but for quadrant 4 (near limb).}
	\label{sim_4_l}
\end{figure}

\end{appendix}

\end{document}